\documentclass[12pt]{article}
\usepackage{amsmath}
\usepackage{amssymb}
\begin{document}
\begin{center}
{\large\bf  {A Quantum Mechanical Example for Hodge Theory }}

\vskip 2.0cm

{\sf Shri Krishna$^{(a)}$, R. P. Malik$^{(b,c)}$}\\

$^{(a)}$ {\it Department of Physics, Zakir Husain Delhi College,}\\
{\it University of Delhi, Delhi -  110002, India}\\

\vskip 0.2cm

$^{(b)}$ {\it Physics Department, Institute of Science,}\\
{\it Banaras Hindu University, Varanasi - 221 005, (U.P.), India}\\

\vskip 0.2cm

\vskip 0.2cm

$^{(c)}$ {\it DST Centre for Interdisciplinary Mathematical Sciences,}\\
{\it Institute of Science, Banaras Hindu University, Varanasi - 221 005, India}\\
{\small {\sf {e-mails: skrishna.bhu@gmail.com; rpmalik1995@gmail.com; malik@bhu.ac.in}}}

\end{center}

\vskip 1cm

\noindent
{\bf Abstract:}
On the basis of (i) the discrete and continuous symmetries (and corresponding conserved charges), (ii) the 
ensuing algebraic structures of the symmetry
operators and conserved charges, and (iii) 
a few basic concepts behind the subject of differential geometry, we show that the celebrated Friedberg-Lee-Pang-Ren
(FLPR) quantum mechanical model (describing the motion of a single non-relativistic particle of unit mass under the influence of a 
general {\it spatial} 2D rotationally invariant potential) provides a tractable physical example for
the Hodge theory within the framework of Becchi-Rouet-Stora-Tyutin (BRST)
formalism where the symmetry operators and conserved charges lead to the physical
realizations of the de Rham cohomological operators of differential geometry at the {\it algebraic} level. We concisely mention the Hodge decomposition theorem in the quantum Hilbert space of states and choose the harmonic states as
 the {\it real} physical states of our theory.
We discuss the physicality criteria w.r.t. the conserved and nilpotent versions of the (anti-)BRST and (anti-)co-BRST charges and 
the physical consequences that ensue from them.  


\newpage

\section{Introduction}

Some of the basic concepts and a few 
key ideas of pure mathematics have been at the heart of the modern developments in the realm of theoretical physics
(at  their highest level of sophistication). In particular, the ideas of the (super)string theories (see, e.g. [1-5] and references therein),
in the domain of theoretical (high energy) physics\footnote{As far as the realm of quantum field theories is concerned, the research activities
in the domains of supersymmetric Yang-Mills theories and topological field theories are a couple of examples where the convergence
of ideas, from the pure mathematics and theoretical physics, is found to be outlandish.}, have brought together the active theoretical physicists and mathematicians on a single intellectual 
platform where {\it both} classes of researchers have acquired deeper understanding of some of the actual physical phenomena of nature. We have devoted
past few years in studying the theoretical aspects of (i) some of the field-theoretic models 
of Hodge theory in the specific dimensions of the flat Minkowskian spacetime (see. e.g. [6-8] and references therein), (ii) a few key examples
of Hodge theory in the context of a set of interesting and well-known $\mathcal {N} = 2 $ SUSY quantum mechanical models 
(see, e.g. [9-11] and references therein), and (iii) a toy model of Hodge theory as a rigid rotor [12], etc. All these models of
Hodge theory have brought together the basic concepts and ideas behind the mathematics of 
the de Rham cohomological operators of differential geometry [13-17] and some of the  discrete as well as the continuous symmetries
(and corresponding Noether conserved charges) of the theoretical physics behind the 
basic ideas of Becchi-Rouet-Stora-Tyutin (BRST) formalism [18-21]. In these models of Hodge theory,
the {\it algebraic} structures have been given utmost importance (at various levels of mathematics as well as BRST formalism). 
The central purpose of our present investigation is to demonstrate that the
one (0 + 1)-dimensional (1D) quantum mechanical system of the celebrated Friedberg-Lee-Pang-Ren (FLPR) model [22] is an interesting 
example of Hodge theory within the framework of BRST formalism where the symmetries 
and Norther conserved charges play a decisive role.

Against the backdrop of the key contents that are contained in the above paragraph, it is pertinent to point out the 
importance of the FLPR model
in the context of the one (0 + 1) dimensional (1D) gauge theory [22] where the {\it correct} results have been obtained by taking into account
{\it all} the Gribov-type equivalent copies. This theoretical approach
 is {\it different} from the very important insight given by Gribov himself in his seminal
work [23] as far as the issue of the gauge-fixing (and the ambiguity associated with it) in the context of QCD is concerned. The FLPR
model is a quantum mechanical system which happens to be a 1D
reduction of the D-dimensional Yang-Mills theories that (i) describes the motion of a non-relativistic single 
 particle of unit mass under the influence of a general {\it spatial} 2D rotationally invariant potential of the form: $U (x^2 + y^2)$, and (ii)
is endowed with the first-class constraints in the terminology 
of Dirac's prescription for the classification scheme of constraints (see, e.g. [24,25])
 which generate a set of  infinitesimal local gauge symmetry transformations.
Using the path integral approach and taking into account {\it all} the Gribov-type equivalent copies, the BRST analysis of the FLPR
model has been considered in [26]. It is clear from the above discussions that the root-cause behind the Gribov ambiguity is the 
gauge-fixing term (which is invoked for the quantization purposes). This gauge-fixing term {\it itself} has been avoided in a recent work [27]
on the quantization scheme of the FLPR model
which is based on the physical projection operator(s) that have been proposed in an earlier work [28] for 
the quantization of a few  models of physical interest. In a recent work [29], the {\it modified} Faddeev-Jackiw technique [30,31] has
been used for the quantization of {\it this}  model where the gauge symmetry transformations and the corresponding (anti-)BRST symmetry transformations
have {\it also} been pointed out in the Cartesian as well as the polar coordinate systems by taking into account an admissible gauge-fixing term
[cf. Eq. (1) and footnote after Eq. (2) below].
We have been able to demonstrate the existence of the off-shell nilpotent and absolutely anticommuting
(anti-)co-BRST symmetry transformations [in addition to the {\it usual} (anti-)BRST
symmetry transformations] for {\it this} model in the Cartesian coordinate system in our recent work [32]. In a very recent work [33], 
this model has been {\it also} shown to be an example of the Hodge theory in the {\it polar} coordinate system where {\it only} the continuous 
symmetry transformations have been discussed (and the nilpotent symmetries have been derived by applying the supervariable approach).

In our present endeavor, we demonstrate the existence of {\it six} continuous and {\it a couple} of discrete duality symmetry transformations
(within the framework of BRST formalism)  for the 
1D quantum mechanical system of the FLPR model in the Cartesian coordinate system to establish that this 
physically interesting model is a 1D quantum  mechanical example for the Hodge theory where the discrete and continuous symmetry transformations
(and corresponding conserved charges) provide the physical realizations of the de Rham cohomoloical operators of differential
geometry [13-17] at the {\it algebraic} level. Out of the above six continuous symmetry transformations, {\it four} are fermionic
(i.e. off-shell nilpotent of order two and absolutely anticommuting) in nature and 
the rest of the {\it two} are bosonic. The existence of
a couple of discrete duality symmetry transformations  is very crucial as the interplay between {\it these} and the off-shell nilpotent
(i.e. fermionic) symmetry transformations provide the physical realizations [cf. Eq. (28) below] of the mathematical 
relationship (i.e. $\delta = \pm\, *\, d\, * $) that exists between the nilpotent ($\delta^2 = d^2 = 0$) (co-)exterior
derivatives $(\delta)d$ of differential geometry [13-17]. The total set of discrete duality symmetry transformations 
provide the physical realization of the Hodge duality $*$ operation of the differential geometry that is present in the
relationship $\delta = \pm\, *\, d\, * $. We briefly mention, in our present endeavor, the celebrated Hodge decomposition theorem
on the physical states (i.e. $|phys>$), in the total quantum Hilbert space of states, and discuss the physicality criteria (cf. Sec. 6 below)
w.r.t. the off-shell nilpotent and conserved (anti-)BRST as well as the (anti-)co-BRST charges by choosing the {\it harmonic}
states as the {\it real} physical states (i.e. $|Rphys>$) and comment on the physical consequences that ensue
from the physicality criteria where we highlight the importance of the Dirac quantization conditions on the
real physical states (i.e. $|Rphys>$) in the language of the operator form of the first-class constraints of the {\it original}
classical theory that is described by the first-order Lagrangian $L_f$ [cf. Eq. (1) below]. To be more precise, we demonstrate that the
operator form of the first-class constraints annihilate the real physical states at the {\it quantum} level.

The following key motivating factors have spurred our interest in our present investigation. First of all, in a very recent work [32],
we have been able to show the existence of the off-shell nilpotent and absolutely anticommuting (anti-)co-BRST symmetry
transformations over and above the fermionic (anti-)BRST symmetry transformations that have been pointed out and discussed  in [29]
for the quantum mechanical one (0 + 1)-dimensional (1D) FLPR model. The
existence of the above  {\it four} fermionic
asymmetry transformations are the most fundamental symmetries at the heart of the proof of a theory to be a tractable
model for the Hodge theory. Second, we have focused our discussion on the (anti-)BRST and (anti-)co-BRST invariant first-order form of the
quantum Lagrangian for the above FLPR model in the context of our threadbare discussions because it contains more number of variables
than its counterpart second-order Lagrangian. Hence, as far as the theoretical discussions are concerned, the {\it former} Lagrangian 
is more appealing and interesting to us (see, e.g. [34]). Third, if the beauty, strength and importance of a theory
is defined in terms of the number and variety of symmetries it respects, the 
D-dimensional ($D \geq 2$) field-theoretic as well as the one dimensional (1D) toy models of Hodge theories
belong to this class. In our present endeavor, we demonstrate the existence of a set of {\it six} continuous and a couple of discrete
symmetry transformations for the 1D quantum mechanical FLPR model 
of a gauge theory (where the {\it correct} results have been obtained by taking into account {\it all}
the Gribov-type equivalent copies). Finally, the 1D quantum mechanical model of FLPR is a very interesting model
for a gauge theory which should be explored from many different angles. Our present endeavor is a modest attempt in that direction
where a great deal of emphasis is laid on the symmetry properties and ensuing conserved charges.

The theoretical materials of our present investigations are organized as follows. In Sec. 2, 
we recapitulate the bare essentials of the off-shell nilpotent and absolutely anticommuting
(anti-)BRST and (anti-)co-BRST symmetry transformations and their generators as the 
conserved and off-shell nilpotent (anti-)BRST and (anti-)co-BRST charges, respectively. Our Sec. 3
is devoted to an elaborate discussion on a {\it unique} bosonic transformation
that emerges out from the anticommutator of the above nilpotent (anti-)BRST and (anti-) co-BRST 
transformations. In our Sec. 4, we devote time on the existence of the
ghost-scale symmetry transformations and derive their generator as the conserved 
ghost charge. This charge satisfies the extended  BRST algebra with the (anti-)BRST, (anti-)co-BRST
and a {\it unique} bosonic charge which is also reflected in the algebra that is
satisfied by the operator form of these symmetry transformations. Our Sec. 5 deals with the existence of a couple of discrete duality
symmetry transformations that establish a couple of very specific connections between the nilpotent
(anti-)BRST and (anti-)co-BRST symmetry transformations. The theoretical material of our Sec. 6 concerns itself with
the brief discussion on the Hodge decomposition theorem in the quantum Hilbert space of states and the physicality criteria
w.r.t. the nilpotent and conserved (anti-)BRST and (anti-)co-BRST charges.
Finally, in Sec. 7, we summarize our key results and comment
on the physical realizations of the de Rham cohomological operators of differential geometry at the {\it algebraic} level
where the symmetry operators and the conserved charges play the decisive roles.

In our Appendices A, B and C, we collect some explicit computations 
that are useful in the understanding 
of a few key issues and equations in the main body of our text. \\

\noindent
{\it Conventions and Notations:} We adopt the cnvention of the {\it left} derivative w.r.t. all the {\it fermionic} variables of
our theory in the computations of the EL-EoMs, conjugate momenta, Noether conserved conserved  charges, etc. In the whole body of our text, we denote the 
nilpotent (anti-)BRST and (anti-)co-BRST transformations by the symbols $s_{(a)b}$ and $s_{(a)d}$, respctively,
which anticommute with all the fermionic variables and commute with all the bosonic variables of our theory. The notations
$Q_{ (a)b}$ and $Q_{(a)d}$ stand for the corresponding conserved nilpotent 
charges, respectively. To maintain the consistent and systematic notations, {\it even} the infinitesimal bosonic symmetry transformations
(e.g. a unique bosonic and the ghost-scale symmetries) have been denoted by the symbols: $ (s_\omega, \,s_{\bar\omega})$ and  $s_g$, respectively.
However, it is pretty obvious that $s_{\omega}^2 \neq 0, \, s_{\bar \omega}^2 \neq 0$ and $s_g^2 \neq 0$.\\

\section{Preliminaries: Nilpotent Symmetries}

Our present section is divided into two parts. In subsection 2.1, we discuss very concisely a few aspects of the (anti-)BRST symmetry
transformations and corresponding conserved charges. Our subsection 2.2 is devoted to a brief description of the (anti-)co-BRST symmetry 
transformations and the ensuing conserved charges.

\subsection{ (Anti-)BRST Symmetries: Key Aspects}

We begin with the (anti-)BRST invariant {\it quantum} version of the Lagrangian $L_b$ that includes the {\it classical} first-order Lagrangian
$L_f$ for the FLPR model and incorporates into itself the gauge-fixing and Faddeev-Popov (FP) ghost terms as (see. e.g. [32,29] for details)
\begin{eqnarray}
L_{b} &=& L_f -\; \frac{1}{2}\, (\dot \zeta - z)^2  - i \, \dot {\bar c}\, \dot c -\, i\,\bar c\, c, \nonumber\\
&\equiv& p_x \, \dot x + p_y \, \dot y + p_z \, \dot z - \frac{1}{2}\, \big (p_x^2 + p_y^2 + p_z^2 \big )
-\zeta\, \bigl [g \,(x\,p_y - y\, p_x) + p_z \bigr ] - U(x^2 + y^2) \nonumber\\
&+& b \,\bigl (\dot \zeta - z \bigr ) + \frac{1}{2}\, b^2 - i \, \dot {\bar c}\, \dot c -\, i\,\bar c\; c,
\end{eqnarray}
where, as is obvious, the {\it classical} first-order Lagrangian is
\begin{eqnarray}
L_{f} = p_x \, \dot x + p_y \, \dot y + p_z \, \dot z - \frac{1}{2}\, \big (p_x^2 + p_y^2 + p_z^2 \big )
-\zeta \, \bigl [g \, (x\,p_y - y\, p_x) + p_z \bigr ] - U(x^2 + y^2),
\end{eqnarray}
with a general {\it spatial} 2D rotationally invariant potential function of the form: $ U(x^2 + y^2)$. 
In the above, the set $(\dot x, \dot y, \dot z)$ are the generalized velocities 
[with $\dot x = (dx/dt), \dot y = (dy/dt), \dot z = (dz/dt)$] and the set ($p_x,\; p_y,\; p_z$) are the canonical conjugate momenta
corresponding to the set of Cartesian coordinates $(x, y, z)$. 
Here the evolution parameter is the ``time'' variable $t$ which varies from $-\, \infty $ to $+\, \infty $. 
The Lagrange multiplier variable $\zeta(t)$ behaves like the
``gauge'' variable of our theory and the ferminonic ($c^2 = \bar c^2 = 0, \; c\, \bar c + \bar c\, c = 0$) (anti-)ghost
variables $(\bar c)c$ are required for the validity of 
the unitarity in our theory. The Nakanishi-Lautrup type auxiliary variable $b$ has been
invoked for the linearization of the {\it quadratic} 
gauge-fixing\footnote{This gauge-fixing term is like the famous 't Hooft gauge-fixing term (see, e.g. [35,36]) in the context of
the St$\ddot u$ckelberg-modified Proca theory which is described by the Lagrangian density:$-\, \frac{1}{4} F^{\mu\nu}\, F_{\mu\nu}
+ \frac{1}{2}\; m^2\,A^\mu\, A_\mu \mp\, m\, A^\mu\, \partial_\mu \phi + \frac{1}{2}\, \partial^\mu \, \phi\, \partial_\mu \, \phi
-\; \frac{1}{2} \, (\partial_\mu A^\mu \pm m\, \phi)^2$ where the St$\ddot u$ckelberg-modified Abelian 
massive vector boson is modified as: $A_\mu \to A_\mu \mp\, \frac{1}{m}\, \partial_\mu\,\phi$ where $\phi$ is a pure scalar field. In
this Lagrangian density $F_{\mu\nu} = \partial_\mu A_\nu - \partial_\nu A_\mu$ is the field strength tensor and $m$ is the rest 
mass of the vector boson $A_\mu$. The gauge-fixing term $- \frac{1}{2}\, (\dot \zeta - z)^2$  in (1) is just like the 't Hooft gauge-fixing term
with the identifications: $\partial_\mu A^\mu \sim \dot \zeta$ and $\phi \sim z$ for our 1D system of the FLPR model (with unit mass).
As we observe in (3) that $s_{(a)b} \, p_z = 0$, similarly we find that $s_{(a)b} \, \Pi_\phi = 0$ (see, e.g. [36]) where 
$\Pi_\phi = \partial_0 \phi \mp m\, A_0$ is the conjugate momentum w.r.t. the $\phi$ field that is derived from the above St$\ddot u$ckelberg-modified
Lagrangian density.} 
term  [i.e. $  b\, (\dot \zeta - z) + (b^2/2) \equiv - (1/2)\, (\dot \zeta - z)^2 $] that is present in the 
quantum version of the (anti-)BRST invariant Lagrangian (1).

The Lagrangian $L_b$ respects the following infinitesimal, continuous, off-shell nilpotent [$s_{(a)b}^2 = 0$] and absolutely
anticommuting [i.e. $(s_b + s_{ab})^2 \equiv s_b \, s_{ab} + s_{ab}\, s_b 
\equiv \{ s_b, \; s_{ab} \} = 0$] (anti-)BRST symmetry transformations [$s_{(a)b}$] (see, e.g. [32] for details)
\begin{eqnarray}
 s_{ab} \, x &=& -\, g\, y \,\bar c, \quad s_{ab}\, y = g \,x \,\bar c, \quad s_{ab}\, z = \bar c, \quad s_{ab}\, \zeta  = \dot {\bar c},
\quad s_{ab}\, p_x = -\, g\, p_y\, \bar c, \nonumber\\
s_{ab}\, p_y &=&  g \,p_x \,\bar c, \quad s_{ab}\, p_z = 0,
\quad s_{ab}\, p_\zeta  = 0,  \quad s_{ab} \,\bar c = 0,
\quad s_{ab}\,c = - i\, b, \quad s_{ab}\,b = 0, \nonumber\\
 s_{b} \, x &=& -\, g\, y \,c, \qquad s_{b}\, y = g \,x \, c, \qquad s_{b}\, z =  c, \qquad s_{b}\, \zeta  = \dot  c,
\qquad s_{b}\, p_x = -\, g\, p_y\,  c, \nonumber\\
s_{b}\, p_y &=&  g \,p_x \, c, \qquad s_{b}\, p_z = 0,
\quad s_{b}\, p_\zeta  = 0,  \quad s_{b} \, c = 0,
\quad s_{b}\,\bar c =  i\, b, \qquad s_{b}\,b = 0,
\end{eqnarray}
because we observe that the quantum Lagrangian $L_b$ transforms 
to the total time derivatives (under the above infinitesimal, continuous and fermionic transformations) as
\begin{eqnarray}
    s_{ab} L_b \,= \,\frac{d}{dt}\, (b \; \dot{\bar c}), \qquad \qquad s_{b} L_b \,= \,\frac{d}{dt}\, (b \; \dot{c}),
\end{eqnarray}
which render the action integral invariant\footnote{In our present investigation, we have concentrated {\it only} on (i) the
symmetry invariance of the action integral, (ii)  the ensuing Noether conserved charges, and (iii) the physicality criteria w.r.t. them
(cf. Sec. 6 below). However, in the realm of the path integral approach to any arbitrary quantum systems, it is the transition
amplitude that plays an important role. The {\it latter} may not remain invariant under various symmetry transformations
of the action integral and, for our 1D quantum system, these symmetries might be spontaneously broken. 
Even though these issues are very important, we have {\it not} devoted time on them 
in our present endeavor. For interested readers' convenience,
we would like to point out that a very good account of the Gribov copying and BRST formalism, various issues connected with
the path integral approach, Batalin-Vilkovisky theorem, supplementary conditions, etc., have been carried out in [37].}
because all the physical variables vanish off as $ t \to \pm \infty$. Hence, the (anti-)BRST transformations (3) are the {\it symmetry}
transformations for the action integral corresponding to the quantum Lagrangian $L_b$. As a consequence, we have the following 
expressions for the conserved (anti-)BRST charges [$Q_{(a)b}$], namely;
\begin{eqnarray}
Q_{ab} &=& \bigl [g \, (x\, p_y - y \, p_x) + p_z \bigr ]\, \bar c + b \, \dot {\bar c}
 \equiv b\, \dot {\bar c} - \dot b \, \bar c, \nonumber\\
Q_{b} &=& \bigl [g \, (x\, p_y - y \, p_x) + p_z \bigr ]\,  c + b \, \dot {c}
 \equiv b\, \dot {c} - \dot b \, c,
\end{eqnarray}
where, to obtain the concise forms of the charges, we have used the equation of motion: $\dot b = - \bigl [g \, (x\, p_y - y \, p_x) + p_z \bigr ]$.
The conservation laws (i.e. $\dot Q_{(a)b} = 0$) for the above charges 
[cf. Eq. (5)] can be proven, in a straightforward manner, by using the
appropriate Euler-Lagrange (EL) equations of motion (EoM) that emerge out from the Lagrangian $L_b$.

We end this subsection with the following key and clinching remarks. First, the conserved charges $Q_{(a)b}$ are also
off-shell nilpotent (i.e. $Q_{(a)b}^2 = 0$) and absolutely anticommuting 
(i.e. $Q_b \, Q_{ab} + Q_{ab} \, Q_b = 0$) in nature (see, e.g. [32] for details). Second, the total kinetic terms of the quantum
version of the Lagrangian $L_b$ remain invariant under the (anti-)BRST symmetry transformations. Third, being fermionic in nature, the (anti-)BRST symmetry 
transformations transform the bosonic variables into fermionic variables and vice-versa. However, these 
fermionic symmetries are {\it not} like the fermionic
${\mathcal N} = 2$ SUSY symmetry transformations where there is {\it no} validity of the absolute anticommutativity property. 
Fourth, it is interesting to point out that the 
classically gauge invariant (see, e.g. [32] for details) potential function $U(x^2 + y^2)$ remains invariant under the nilpotent
(anti-)BRST symmetry transformations (3) at the quantum level, too.
Finally, the (anti-)BRST charges are the generators for the off-shell nilpotent (anti-)BRST transformations (3) because we observe that the 
following relationship is true, namely;
\begin{eqnarray}
s_r\, \phi (t) = -i\, \bigl [\phi (t), \; Q_r \bigr ]_{(\pm)} \qquad \mbox{with} \qquad r = b, ab, 
\end{eqnarray}
where the ${(\pm)}$ signs, as the subscripts on the square bracket, stand for the 
above square bracket to be an (anti)commutator for the  generic variable $\phi (t)$
of the quantum version of the (anti-)BRST invariant Lagrangian $L_b$ being  fermionic (e.g. $\phi = \bar c,\, c $) and  
bosonic (e.g. $\phi = x, y, z, \zeta, p_x, p_y, p_z, p_\zeta, b$)
in nature, respectively.

\subsection{(Anti-)co-BRST Symmetries: Salient Features}

In addition to the off-shell nilpotent and absolutely anticommuting (anti-)BRST symmetry transformations (3), we observe that the
quantum version of the Lagrangian $L_b$ respects (see, e.g. [32] for details) off-shell nilpotent 
(i.e. $s_{(a)d}^2 = 0$) and absolutely anticommuting [i.e. $(s_d + s_{ad})^2 \equiv s_d \, s_{ad} + s_{ad}\, s_d 
\equiv \{ s_d, \; s_{ad} \} = 0$] 
(anti-)co-BRST (i.e. dual-BRST and anti-dual-BRST) symmetry transformations [$s_{(a)d}$], namely;
\begin{eqnarray}
 s_{ad} \, x &=& -\, g\, y \,\dot {c}, \quad s_{ad}\, y = g \,x \,\dot c, \quad s_{ad}\, z = \dot c, \quad s_{ad}\, \zeta  = c,
\quad s_{ad}\, p_x = -\, g\, p_y\, \dot c, \quad s_{ad}\,c = 0, \nonumber\\
s_{ad}\, p_y &=&  g \,p_x \,\dot c, \quad s_{ad}\, p_z = 0,
\quad s_{ad}\, p_\zeta  = 0,  \quad s_{ad} \,\bar c = i\, \bigl [g (x\, p_y - y\, p_x) + p_z \bigr ], \quad s_{ad}\,b = 0,
\nonumber\\
 s_{d} \, x &=& -\, g\, y \,\dot {\bar c}, \quad s_{d}\, y = g \,x \,\dot {\bar c} , \quad s_{d}\, z = \dot {\bar  c},
\quad s_{d}\, \zeta  = \bar c,
\quad s_{d}\, p_x = -\, g\, p_y\, \dot {\bar c}, \quad s_{d}\,b = 0, \nonumber\\
s_{d}\, p_y &=&  g \,p_x \, \dot {\bar c}, \quad s_{d}\, p_z = 0,
\quad s_{d}\, p_\zeta  = 0,  \quad s_{d} \, \bar c = 0,
\quad s_{d}\, c = - \, i\, \bigl [g (x\, p_y - y\, p_x) + p_z \bigr ],
\end{eqnarray}
because of the observations that the quantum Lagrangian $L_b$ transforms to the total time derivatives 
under the above infinitesimal  and  continuous transformations, namely;
\begin{eqnarray}
s_{ad}\, L_b &=& \frac{d}{dt}\, \Bigl [ \{g \, (x\, p_y - y \, p_x) + p_z \}\, \dot c \Bigr ], \nonumber\\
s_{d}\, L_b &=& \frac{d}{dt}\, \Bigl [ \{g \, (x\, p_y - y \, p_x) + p_z \}\, \dot {\bar c} \Bigr ].
\end{eqnarray}
As a consequence, the action integral, corresponding to the Lagrangian $L_b$, remains invariant under the above 
(anti-)co-BRST symmetry transformations (7) because all the physical variables vanish off as $t \to \pm \infty$.
According to the  celebrated Noether theorem, the above observations in (8), lead to the derivations of the 
(anti-)co-BRST charges [$Q_{(a)d}$] as [32]:
 \begin{eqnarray}
Q_{ad} &=& \bigl [g\, (x\, p_y - y \, p_x) + p_z \bigr ] \, \dot c + b\, c \equiv b\, c - \dot b\, \dot c, \nonumber\\
Q_{d} &=& \bigl [g\, (x\, p_y - y \, p_x) + p_z \bigr ] \, \dot {\bar c} + b\, \bar c \equiv b\, {\bar c} - \dot b\, \dot {\bar c}.
\end{eqnarray}
The conservation (i.e. $\dot Q_{(a)d} = 0$) of the above charges can be proven in a straightforward fashion by exploiting the
beauty and strength of the following EL-EoMs:
\begin{eqnarray}
&&\dot x = p_x - g\,\zeta\,y, \qquad \dot y = p_x + g\,\zeta\,x, \qquad \dot z = p_z + \zeta\, \qquad \dot p_z = -\, b, 
\qquad \ddot b = b, \nonumber\\
&&\dot p_x =- g\,\zeta\,p_y -\, 2\,x\, U^\prime, \quad \dot p_y = g\, \zeta\,p_x  -\, 2\,y\, U^\prime, \quad
\dot b = -\, \bigl [g\, (x\, p_y - y \, p_x) + p_z \bigr ], \nonumber\\
&& \ddot c = c, \qquad \ddot {\bar  c} = {\bar c}, \qquad \ddot p_z = \bigl [g\, (x\, p_y - y \, p_x) + p_z \bigr ] \equiv -\, \dot b.
\end{eqnarray}
In the above, the symbol $U^\prime$ stands for the derivative of the rotationally invariant potential $U$ with respect
to its full argument : $x^2 + y^2$. In the verification of the very useful equation of motion: $\ddot b = b$, many other
EL-EoMs (derived from $L_b$)  are to be used. This equation of motion plays a crucial role in the proof of the 
conservation laws for the (anti-)co-BRST charges (9) which are the generators for the (anti-)co-BRST symmetry
transformations (7). These charges can be used in (6) where there is a universal relationship between the continuous
symmetries and their generators as the Noether conserved charges. For this purpose, we have to {\it only} replace in (6) the symbol
$ r = b, ab$ by the substitutions: $ r = d, ad$.

We end this short (but very important) subsection with the following crucial remarks. First, it is very straightforward to verify that the
(anti-)BRST and (anti-)co-BRST symmetry transformation operators satisfy the following very useful anticommutators, namely;
\begin{eqnarray}
\{s_b,\; s_{ab}\} = 0, \qquad \{s_b,\; s_{ad}\} = 0, \qquad \{s_d,\; s_{ad}\} = 0, \qquad \{s_d,\; s_{ab}\} = 0,    
\end{eqnarray}
which establishes the absolute anticommutativity properties amongst {\it some} of the fermionic (anti-)BRST and (anti-)co-BRST symmetry
transformations (that correspond to the infinitesimal, continuous and off-shell nilpotent  transformation operators that have
been listed in (3) and (7), respectively). Second, it is self-evident [in view of our observations in (11)] that, at this juncture, 
out of the {\it four} existing off-shell nilpotent
transformations in the theory, we have only the following {\it two} non-zero anticommutators that correspond to the existence of the
bosonic ($s_\omega^2 \neq 0, \, s_{\bar \omega}^2 \neq 0 $) symmetry transformations ($ s_\omega, \, s_{\bar \omega} $), namely;
\begin{eqnarray}
  s_\omega = \{s_b, \; s_d \}, \qquad \qquad  s_{\bar \omega} = \{s_{ab}, \; s_{ad} \}.
\end{eqnarray}
We shall see, in the forthcoming next section, that out of the above two bosonic symmetry operators, only {\it one} is independent. Hence, 
there is a {\it unique}
bosonic symmetry in our theory. Third, it is straightforward to note that the fermionic (anti-)co-BRST charges [cf. Eq. (9)] are off-shell
nilpotent (i.e. $Q_{(a)d}^2 = 0$) of order two and they are found to be absolutely anticommuting (i.e. $Q_d\, Q_{ad} + Q_{ad}\, Q_d = 0$)
in nature (see, e.g. [32] for details). The {\it latter} property establishes the linear independence of the co-BRST and anti-co-BRST charges.
Finally, we note that one of the key signatures of the (anti-)co-BRST symmetry transformations is the observation
that the {\it total} gauge-fixing terms [i.e. $b \,\bigl (\dot \zeta - z \bigr ) + (1/2)\, b^2$] remain invariant under them. This salient feature
of the (anti-)co-BRST symmetry transformations distinguishes {\it them} from the (anti-)BRST symmetry transformations under which the {\it total}
kinetic terms of the Lagrangian (1) remain invariant (cf. Subsec. 2.1).\\

\section{Bosonic Symmetry Transformations: Uniqueness}

One of the key purposes of this section is to show that we have the validity of  $s_\omega + s_{\bar \omega} = 0$ at {\it several} stages
(in our theoretical discussions) which proves the {\it uniqueness} of the bosonic transformations (12) that have been
defined in terms of the {\it specific} anticommutators between the (anti-)BRST and (anti-)co-BRST  transformations. Under the bosonic
symmetry transformation $s_\omega = \{s_b, \; s_d \}$, we note that the variables of the quantum version of the Lagrangian $L_b$
transform in the following manner (modulo a factor of $i$), namely; 
\begin{eqnarray}
 s_\omega  x &=& -\, g\, y\, (\dot b - \Omega), \; \quad  s_\omega  y = + \,g\, x\, (\dot b - \Omega),  \; \quad s_\omega b = 0, \;
  \quad s_\omega p_\zeta =0, \nonumber\\
 s_\omega  z &=&  +  \, (\dot b - \Omega),\; \; \qquad
 s_\omega  \zeta = + \,  (b - \dot \Omega), \; \qquad s_\omega c = 0, \; \qquad s_\omega \bar c = 0,\nonumber\\
s_\omega  p_x &=& -\, g\, p_y\, (\dot b - \Omega),  \quad  s_\omega  p_y = +\, g\, p_x\, (\dot b - \Omega),  \quad s_\omega  p_z = 0, \quad s_\omega \, \Omega = 0,
  \end{eqnarray}
where we have adopted the notation: $\Omega = g \, (x\, p_y - y \, p_x) + p_z  $ for the sake of brevity. In exactly similar fashion,
we note that {\it all} the variables of the Lagrangian $L_b$ [cf. Eq. (1)], transform when we take into account the definition
of the bosonic symmetry transformation as: $s_{\bar \omega} = \{s_{ab}, \; s_{ad} \}$. In other words, we have the 
following bosonic symmetry transformations under $s_{\bar \omega}$ [that has been defined in (12)], namely;
\begin{eqnarray}
 s_{\bar\omega}  x &=& +\, g\, y\, (\dot b - \Omega), \; \quad  s_{\bar\omega} y = - \,g\, x\, (\dot b - \Omega),  \;\quad s_{\bar\omega} b = 0,
  \; \quad s_{\bar\omega} p_\zeta =0, \nonumber\\
 s_{\bar\omega} z &=&  -  \, (\dot b - \Omega), \;\;  \qquad
 s_{\bar\omega}  \zeta = - \,  (b - \dot \Omega), \;\, \qquad s_{\bar\omega} c = 0, \; \qquad s_{\bar\omega} \bar c = 0,\nonumber\\
s_{\bar\omega} p_x &=& +\, g\, p_y\, (\dot b - \Omega),  \quad  s_{\bar\omega}  p_y = -\, g\, p_x\, (\dot b - \Omega),  \quad s_{\bar\omega}  p_z = 0,
\quad s_{\bar\omega} \, \Omega = 0,
  \end{eqnarray}
modulo a factor of $i$. A close look at (13) and (14) demonstrate, in a very straightforward manner, that we have a {\it unique} 
bosonic symmetry transformation
in our theory (which is defined in terms of the 
{\it specific} anticommutators [cf. Eq. (12)] of the off-shell nilpotent and absolutely anticommuting (anti-)BRST and (anti-)co-BRST
symmetry transformations). This is because 
of the fact that the operator form of the transformations $s_\omega$ and $s_{\bar \omega}$ satisfy the following straightforward relationship:
\begin{eqnarray}
  s_\omega \,+ \, s_{\bar \omega} = 0.
\end{eqnarray}
The above observation (and the related statement) is {\it true} even at the level of symmetry transformation of the (anti-)BRST and (anti-)co-BRST invariant
Lagrangian $L_b$, too.

To corroborate the above claim as well as the statement, we concentrate on the 
bosonic symmetry transformations of the quantum version of
the (anti-)BRST as well as the (anti-) co-BRST invariant Lagrangian $L_b$. In this context,
we find that the following transformations on the Lagrangian $L_b$ are {\it correct}, namely;
\begin{eqnarray}
    s_\omega L_b &=& \frac{d}{dt} \, \Bigl [ \dot b\, \bigl \{g\,(x\, p_y - y \,p_x) + p_z \bigr \}\, 
    -  b \, \frac{d}{dt}\, \bigl \{ g\,(x\, p_y - y \,p_x) + p_z \bigr \}  \Bigr ], \nonumber\\
    s_{\bar \omega} L_b &=& \, -\,\frac{d}{dt} \, \Bigl [ \dot b\, \bigl \{g\,(x\, p_y - y \,p_x) + p_z\bigr \}\, 
    - b \, \frac{d}{dt}\, \bigl \{ g\,(x\, p_y - y \,p_x) + p_z \bigr \}  \Bigr ],
\end{eqnarray}    
which establish, once again, that we have the validity of $s_\omega \,+ \, s_{\bar \omega} = 0$ in our theory. The above expressions 
have been obtained\footnote{ There is a simpler way to derive the key results, quoted in (16), in a precise manner. This is due to our 
definitions of the bosonic symmetry operators in (12) and our key observations of the symmetry invariance(s) in (4) and (8).
The direct applications of (12) on the quantum Lagrangian $L_b$ leads to (16).}
after the applications of the symmetry transformations (13) and (14) on every individual term of the full expression for the 
Lagrangian $L_b$ of our theory. The observations in (16),
according to the Noether theorem, lead to the definitions and derivations 
of the bosonic charges (cf. Appendix A for details) as:
\begin{eqnarray}
 Q_\omega &=& b^2 - \bigl [g\,(x\, p_y - y \,p_x) + p_z\bigr ]^2, \nonumber\\
Q_{\bar \omega} &=& \, \bigl [g\,(x\, p_y - y \,p_x) + p_z\bigr ]^2 - b^2.
\end{eqnarray}
As expected, we observe that the above charges are {\it not} independent of each-other. We can obtain the concise forms of the above charges 
by using the EL-EoM w.r.t. the gauge variable $\zeta (t)$ from $L_b$ (i.e. $\dot b = - [g\,(x\, p_y - y \,p_x) + p_z]$)
and $\ddot b = b $. In other words, we have: 
\begin{eqnarray}
    Q_\omega = b^2 - {\dot b}^2 \equiv b\, \ddot{b} - \dot b^2, \qquad \quad
Q_{\bar \omega} = {\dot b}^2 - b^2 \equiv {\dot b}^2 - b\, \ddot b.
\end{eqnarray}
It can be checked that the above charges in (17) and (18) are conserved and they are the generators for the bosonic
symmetry transformations in (13) and (14), respectively. There are a few very simple ways to derive the 
independent version of the bosonic charge $Q_\omega$ that have been discussed, in an elaborate manner, in our Appendix A
where the general (and the standard) relationship between the continuous symmetry transformations and their generators as the conserved 
Noether charges has been taken into account.

We conclude our present section with the following final remarks. First of all, due to our observations in (13) and (14)
that $ s_\omega c = 0, \, s_\omega \bar c = 0, \, s_{\bar \omega} c = 0, s_{\bar \omega} \bar c = 0$, it is self-evident that,
one of the characteristic features of the {\it unique} bosonic symmetry transformations is the observation that the FP-ghost terms
of the Lagrangian $L_b$ remain invariant (i.e. $s_{\omega(\bar \omega)} \, [- i \, \dot {\bar c}\, \dot c -\, i\,\bar c\, c ] = 0$). 
Second, being the anticommutator
(i.e.  $s_\omega = \{s_b, \; s_d \} \equiv -\, \{s_{ab}, \; s_{ad} \}$)
of the fermionic  (anti-)BRST and (anti-)co-BRST symmetry transformations], the {\it unique} bosonic symmetry transformation $s_\omega$
commutes with all the {\it four} off-shell nilpotent symmetry transformations {\it individually}. Thus, it mimics the
algebraic property [cf. Appendix C, Eq. (C.1)]
of the Laplacian operator (present in the set of {\it three} de Rham cohomological operators) of differentials geometry. 
Finally, despite the presence of (i) the {\it four} off-shell nilpotent (i.e. fermionic) (anti-)BRST and (anti-)co-BRST symmetry
transformations [cf. Eqs. (13), (14)], and (ii) the definitions of the {\it two} bosonic symmetry operators in (12), we observe that
only a unique bosonic symmetry transformation $s_\omega $ exists in our 
present 1D quantum mechanical FLPR theory.\\

\section{Ghost-Scale Transformations and Ghost Charge}

In addition to the {\it four} fermionic (anti-)BRST and (anti-)co-BRST symmetry transformations 
[cf. Eqs. (3),(7)] and a {\it unique} set of bosonic
symmetry transformations [see, e.g. Eq. (13)], we observe that our quantum version of Lagrangian
$L_b$ respects the following spacetime independent scale symmetry transformations
\begin{eqnarray}
{\bar  c} \rightarrow e^{-\,\Sigma} \, {\bar c}, \qquad  \quad   c  \rightarrow e^{+\,\Sigma} \, c, 
\qquad \quad \phi \rightarrow e^{0}\, \phi,  
\end{eqnarray}
where $\Sigma$ is the spacetime independent scale transformation parameter. The numerals in the exponents 
of the above scale transformations denote the
ghost numbers $(-1)+1$ for the (anti-)ghost variables $(\bar c)c$, respectively,  and the ghost number {\it zero} is associated with the 
rest of the generic variable $\phi \equiv x, \, p_x,  y, \, p_y,\, z, \, p_z, \zeta, \, p_\zeta, b$ of our theory.
The infinitesimal  versions of the above transformations (19), denoted by the symbol $s_g$, are 
\begin{eqnarray}
 s_g {\bar c} = - \, \bar c, \qquad  \quad  s_g c = + \, c,     \qquad  \quad s_g \phi = 0,
\end{eqnarray}
where, for the sake of brevity, we have chosen the spacetime independent scale transformation parameter to be one (i.e. $\Sigma = + \,1$).
According to the celebrated Noether theorem, the invariance of the quantum version of the Lagrangian $L_b$, 
under the infinitesimal ghost transformations (20), leads to the
following expression for the conserved ghost charge ($Q_g$)
\begin{eqnarray}
 Q_g = i\, \bigl (\bar c\, \dot c - \dot {\bar c}\, c \bigr ),
\end{eqnarray}
whose conservation law (i.e. $\dot Q_g = 0$) can be proven readily by taking the help of the EL-EoMs: $\ddot c = c, \; \ddot {\bar c} = {\bar c}$.
It is elementary to point out that the above conserved ghost charge turns out to be the generator for the infinitesimal ghost scale symmetry
transformations in (20). It can be checked that (i) the FP-ghost part of the Lagrangian $L_b$ remains invariant under the discrete symmetry
transformations\footnote{There is {\it another} set of discrete symmetry transformations in the ghost-sector where we have the transformations:
$c \to \pm\, i\, \dot {\bar c}, \; \bar c \to \pm\, i\, \dot {c}, \; \dot  c \to \mp\, i\,  {\bar c}, \; \dot {\bar c} \to \mp\, i\,  {c}\,$
under which the FP-ghost part (i.e. $- i \, \dot {\bar c}\, \dot c -\, i\,\bar c\; c$)
of the Lagrangian $L_b$ [cf. Eq. (1)] remains invariant. However, there are a couple of key differences between the two. First of all, we note that 
the ghost conserved charge $Q_g$ transforms,
under the above discrete symmetry transformations, as: $Q_g \to -\, Q_g $. Second, we observe that under the former 
(i.e. $c \to \pm i\, \bar c, \; {\bar c} \to \pm i\,  c, \; \dot c \to \mp i\, \dot {\bar c},\; \dot {\bar c} \to \mp i\,  \dot c$) discrete symmetry  transformations, we obtain $- i \, \dot {\bar c}\, \dot c \to - i \, \dot {\bar c}\, \dot c$ and $-\, i\,\bar c\, c \to -\, i\,\bar c \, c$. On the other 
hand, under the {\it alternative} (i.e. the latter) discrete symmetry transformations, we obtain exchange 
relationship: $- i \, \dot {\bar c}\, \dot c  \Leftrightarrow - i \, \dot {\bar c}\, \dot c$ between the two terms of 
the  FP-ghost part of $L_b$. This key observation
is important as it would play an important role in our Appendix B where we shall discuss the duality
symmetry  invariance of the algebra (23) that exist between the conserved ghost charge $Q_g$ and 
the {\it rest} of the conserved charges of our BRST-quantized quantum mechanical system that is described by 
the (anti-)BRST and (anti-)co-BRST invariant Lagrangian $L_b$.}: $ c \to \pm i\, \bar c, \; 
{\bar c} \to \pm i\,  c, \; \dot c \to \mp i\, \dot {\bar c},\; \dot {\bar c} \to \mp i\,  \dot c$, and (ii) {\it these} 
are also the symmetry transformations for the ghost charge $Q_g$ because 
we observe that: $Q_g \to Q_g$ under the above discrete transformations for the (anti-)ghost variables. We shall
see the importance of these discrete symmetry transformations in our next section where we shall elaborate on the
presence of a set of such discrete symmetry transformations for our {\it entire} theory that is described by the 
Lagrangian $L_b$ [cf. Eq. (1)].

Exploiting the theoretical beauty and strength of the relationships between the continuous symmetry transformations and their generators
as the conserved Noether charges, we note that the following standard algebra is satisfied between the ghost charge
and the rest of the other conserved charges of our theory, namely;
\begin{eqnarray}
&& s_g Q_b = -\, i\, \bigl [ Q_b, \; Q_g \bigr ] = +\; Q_b, \qquad  \quad  s_g Q_{ab} = -\, i\; \bigl [ Q_{ab}, \; Q_g \bigr ]  = -\, Q_{ab}, \nonumber\\
&& s_g Q_d = -\, i\, \bigl [ Q_d, \; Q_g \bigr ]  = -\; Q_d, \qquad \quad s_g Q_{ad} = -\,i\, \bigl [Q_{ad}, \; Q_g \bigr ]  = +\; Q_{ad},\nonumber\\
&& s_g Q_\omega  = -\, i\; \bigl [ Q_\omega, \; Q_g \bigr ]  = 0, \qquad \qquad \; \, s_g Q_g = -\, i\, \bigl [ Q_g, \; Q_g \bigr ] = 0,
\end{eqnarray}
where the l.h.s. of the above interesting relationships can be computed explicitly by the applications  of the infinitesimal version of the
ghost scale symmetry transformations (20) on (i) the conserved (anti-)BRST charges [cf. Eq. (5)], (ii) the conserved (anti-)co-BRST charges [cf. Eq. (9)],
(iii) the conserved and unique bosonic charge [cf., e.g., Eq. (17)], and (iv) the conserved ghost charge [cf. Eq. (21)]. Thus, we obtain, ultimately, the following
very beautiful and interesting algebra between the conserved ghost charge and the rest of the other conserved charges of our theory, namely;
\begin{eqnarray}
&&  i\, \bigl [ Q_g, \; Q_b \bigr ] = +\; Q_b, \qquad  \quad   i\, \bigl [ Q_g, \; Q_{ab} \bigr ]  = -\; Q_{ab}, 
\qquad \quad i\, \bigl [ Q_g, \; Q_\omega \bigr ]  = 0,\nonumber\\
&&  i\, \bigl [ Q_g, \; Q_d \bigr ]  = -\; Q_d, \qquad \quad i\, \bigl [Q_{g}, \; Q_{ad} \bigr ]  = +\; Q_{ad}, \qquad \quad
 i\, \bigl [ Q_g, \; Q_g \bigr ] = 0,
 \end{eqnarray}
 which plays a very decisive role in the determination of the ghost numbers of the states in the total
 quantum Hilbert space of states. For instance, let us take the ghost number of a very specific state (i.e. $|\psi>_n$),
 existing in the total quantum Hilbert space of states, equal to  $n$. This statement is defined mathematically through the
 following equation\footnote{It is very clear that, for our theory, the (anti-)ghost variables are $(\bar c)c$ which carry the ghost
 numbers $(-1)+1$, respectively. On the other hand, the {\it basic} physical variables (e.g. $x, \, p_x,  y, \, p_y,\, z, \, p_z, \zeta, \, p_\zeta$)
 of our theory are endowed with the ghost number equal to {\it zero}. Thus, for our specific quantum theory,  the allowed values of 
the ghost numbers are {\it only}: $ n = 0, \, \pm 1$ in our equation (24).}
\begin{eqnarray}
i\; Q_g \, |\psi>_n \,= \,n\; |\psi>_n \qquad \qquad \mbox{with} \qquad \qquad n = 0,\, \pm 1, \,\pm2,...,
\end{eqnarray}
where, as is obvious, the conserved charge $Q_g$ is nothing but the ghost charge [cf. Eq. (21)]. From the above algebraic structure [that is 
present in (23)], it is
straightforward to note that we have the following very interesting relationships
\begin{eqnarray}
&& i\, Q_g \, Q_b \, |\psi>_n \,=\, (n + 1) \;Q_b\, |\psi>_n, \quad    
\qquad  i\, Q_g \, Q_{ab} \, |\psi>_n \,=\, (n - 1) \;Q_{ab}\, |\psi>_n, \nonumber\\
&& i\, Q_g \, Q_{ad} \, |\psi>_n \,=\, (n + 1) \;Q_{ad}\, |\psi>_n, \qquad
 i\, Q_g \, Q_d \, |\psi>_n \,=\, (n - 1) \;Q_d\, |\psi>_n, \nonumber\\
&& i\, Q_g \, Q_\omega \, |\psi>_n \,=\, (n + 0) \;Q_\omega\, |\psi>_n,   
\end{eqnarray}
which demonstrate that the 
specific states: $Q_b \, |\psi>_n, \; Q_{d} \, |\psi>_n $ and $Q_\omega \, |\psi>_n$ carry the ghost numbers
$(n + 1), \; (n -1)$ and $n$, respectively. On the other hand, in exactly similar fashion, the ghost numbers $(n + 1), \; (n -1), \; n$
are associated with the very specific quantum states: $Q_{ad} \, |\psi>_n, \; Q_{ab} \, |\psi>_n, \; Q_\omega \, |\psi>_n$ in the total
quantum Hilbert space of states of our BRST-quantized quantum mechanical theory. We shall see the importance of these
observations in our section on the physicality criteria (cf. Sec. 6 below) where the Hodge decomposition theorem would be
performed in the total quantum Hilbert space of states.

We wrap-up our present section with the following remarks. First of all, we note that (in the total quantum Hilbert space of states),  
there are {\it two} states (i) with the ghost number equal to $(n + 1)$ (e.g. the quantum states: $Q_b \, |\psi>_n, \; Q_{ad} \, |\psi>_n $), and 
(ii)  with the ghost number equal to $(n - 1)$ (e.g. the quantum states: $Q_{ab} \, |\psi>_n, \; Q_{d} \, |\psi>_n $). Second, there is a single
state with the ghost number equal to $n$ (e.g. the quantum state: $Q_\omega \, |\psi>_n$) 
where an operator (e.g. $Q_\omega$) is present [besides the {\it original} state $|\psi>_n$ in (24)
that is present in the definition of the ghost number $n$ itself in 
the language of the conserved ghost charge $Q_g$]. Finally, it can be checked that the algebraic relationships (23) 
amongst the conserved charges are
reflected at the level of the symmetry operators as illustrated below
\begin{eqnarray}
&&  \bigl [ s_g, \; s_b \bigr ] = +\, s_b, \qquad  \quad  \bigl  [ s_g, \; s_{ab} \bigr ]  = -\, s_{ab}, 
\qquad \quad  \bigl [ s_g, \; s_\omega \bigr ]  = 0,\nonumber\\
&&   \bigl [ s_g, \; s_d \bigr ]  = -\, s_d, \qquad \quad  \bigl [s_{g}, \; s_{ad} \bigr ]  = +\, s_{ad}, \qquad \quad
 \bigl [ s_g, \; s_g \bigr ] = 0,
 \end{eqnarray}  
where, as is obvious, the last entries in (23) as well as in (26) are {\it trivially} satisfied.
The sanctity of the above symmetry operator relationships can be checked by applying (i) the above commutators of the symmetry
operators of the l.h.s., and (ii) the resulting symmetry operators of the r.h.s., 
on the generic variable $\phi = x, \, p_x,  y, \, p_y,\, z, \, p_z, \zeta, \, p_\zeta, b$  and the (anti-)ghost variables $(\bar c)c$ of our 
BRST-quantized theory (that is described by the (anti-) BRST and (anti-)co-BRST
invariant quantum version of the Lagrangian $L_b$ [cf. Eq. (1)]).\\

\section{Discrete Duality Transformations: Useful Algebraic Connections Amongst the Nilpotent Symmetries }

Before we come to the main topic of discussion of this section, we would like to pinpoint a few theoretical inputs
that have been given utmost importance in our present section. First of all, we would like to stress that the (anti-)BRST and (anti-)co-BRST
invariant Lagrangian $L_b$ [cf. Eq. (1)] is at the {\it quantum} level. Hence, all the variables 
of this Lagrangian are taken to be {\it hermitian}. Second, all the variables and their first-order time
derivatives are treated as {\it independent} quantities (because we are dealing with the Lagrangian formulation)
and their discrete symmetry transformations will be {\it independent} of one-another. This is clear  from our discussions in
Sec. 4 as well as in the Appendix B where we have taken: $ c \rightarrow 
\; \pm \, i \, \dot {\bar c}, \;\bar c   \rightarrow  \pm \,  i \, \dot {c}, \; \dot c  \rightarrow  \mp \, i \, {\bar c}, \;
\dot{\bar c}  \rightarrow  \mp  \, i \, {c}, \; b  \rightarrow  \pm \, i \, b, \; 
 \dot b \rightarrow   \mp  \, i \, \dot b$. Finally, all the
square terms of the Lagrangian $L_b$ would be taken as: $x^2 = x^\dagger\, x, \; y^2 = y^\dagger\, y, \;
z^2 = z^\dagger\, z, \; b^2 = b^\dagger\, b, \; p_x^2 = p_x^\dagger\, p_x, \; p_y^2 = p_y^\dagger\, p_y, \;
p_z^2 = p_z^\dagger\, p_z $ for the preciseness as well as for the algebraic convenience. Since our system
under discussion is a one (0 + 1)-dimensional (1D) FLPR model, there are {\it no} fundamental matrix representations for
any variable which is treated as an operator. Hence, the dagger operation on a variable is
equivalent to its complex conjugation operation (e.g. $x^\dagger = x^*, \; b^\dagger = b^*$, etc.).
It is straightforward (taking into account the above inputs) that the Lagrangian $L_b$
remains invariant  (i.e. $ L_b \to L_b$) under the
following set of discrete {\it duality} symmetry transformations, namely;
\begin{eqnarray}
&& x \;  \longrightarrow  \; \mp \, i\, x,  \,\;\;\qquad \quad y \; \longrightarrow \; \mp \, i \,y, \,\;\;\qquad \quad
 z \; \longrightarrow \; \mp \, i \, z, \nonumber\\
 && \dot x \;  \longrightarrow  \; \pm\, i\, \dot x, \,\;\;
 \qquad \quad \dot y \; \longrightarrow \; \pm \, i \, \dot y, \;\;\qquad \quad
 \dot z \; \longrightarrow \; \pm \, i \,\dot  z, \nonumber\\
&& p_x \; \longrightarrow \; \mp\, i \, p_x,  \qquad \quad p_y \; 
 \longrightarrow \; \mp \, i \, p_y, \qquad \quad
 p_z \;   \longrightarrow  \; \mp\,   i \, p_z, \; \nonumber\\
 && \dot p_x \; \longrightarrow \; \pm\, i \, \dot p_x, 
\qquad \quad \dot p_y \;  \longrightarrow \; \pm \, i \, \dot p_y,  \qquad \quad
 \dot p_z \;   \longrightarrow  \; \pm\,   i \, \dot p_z, \; \nonumber\\
 && \zeta \; \longrightarrow \; \pm \, i \, \zeta, \;\;\;\qquad \quad
 \dot\zeta \; \longrightarrow \; \mp \, i \, \dot\zeta, \;\;\;\qquad \quad
 g \; \longrightarrow \; \pm \, i \, g, \nonumber\\
 && b \; \longrightarrow \; \pm \, i \, b, \;\;\; \;\qquad \quad 
 \dot b \; \longrightarrow  \; \mp  \, i \, \dot b, \qquad \quad \;\;\;\;
   c \; \longrightarrow 
\; \pm \, i \, \dot {\bar c}, \nonumber\\
&& \bar c \;  \longrightarrow 
\; \pm \,  i \, \dot {c}, \qquad \quad \;\;\;\;
\dot c \; \longrightarrow \; \mp \, i \, {\bar c}, \;\;\;\;
\qquad \quad \dot{\bar c} \; \longrightarrow 
\; \mp  \, i \, {c}. 
\end{eqnarray}
We christen the above discrete set of transformations as the {\it duality} symmetry transformations
because the (anti-)BRST and (anti-)co-BRST charges transform amongst themselves like the electromagnetic
duality transformations that exist between the electric and magnetic fields (cf. Appendix B)
for the invariance of the source-free Maxwell's equations. Some of the other salient features of the above
set of discrete symmetry transformations are the observations that (i) all the square terms of the Lagrangian $L_b$
remain invariant, (ii) the general {\it spatial} 2D rotationally invariant potential function $U(x^2 + y^2)$ transforms to itself
[i.e. $U(x^2 + y^2) \to U(x^2 + y^2$ ], (iii) the FP-ghost terms exchange with each-other
(i.e. $-i\, \dot {\bar c} \, \dot c \Leftrightarrow -i\,  {\bar c} \, c $), and (iv) the kinetic and gauge-fixing terms
remain invariant on their own (i.e. kinetic terms $\to$ kinetic terms,$~$ gauge-fixing terms $\to$ gauge-fixing terms).

At this juncture, it is very interesting and informative to point out that the interplay between the infinitesimal, continuous
and off-shell nilpotent (i.e. $s_{(s)b}^2 = 0, \; s_{(s)d}^2 = 0$) versions of the (anti-)BRST and (anti-)co-BRST symmetry
transformations  and the discrete set of duality symmetry transformations, listed in (27), provide  the 
physical realizations of the mathematical relationship ($\delta = \pm\, *\, d\, *$) that exists between the (co-)exterior derivatives
$(\delta)d$ of the differential geometry [13-17] as follows
\begin{eqnarray}
s_{(a)d}  \,  =  + \; * \, s_{(a)b} \; *,
\qquad \qquad s_{(a)b}  \,  =  - \; * \, s_{(a)d}\; * , 
\end{eqnarray}
where the symbol $*$, in the above equation, corresponds to the discrete duality symmetry transformations (27). In
the proofs of the above relationships, in very special cases, the equations of motions: $\ddot b = b, \;
\dot b = \;- \, \bigl [g\, (x\, p_y - y\, p_x) + p_z \bigr ], \; \ddot c = c, \; \ddot {\bar c} = \bar c$, etc., are invoked. For instance, in the
proofs of $s_b\, \bar c =  -\; *\; s_d\; *\, \bar c$ and $s_d \; c = +\; *\; s_b\; *\, c$, it is very 
important to use the above EL-EoMs. To corroborate these claims, we dwell a bit on the
explicit computations (and their final verifications). Let us focus, first of all, on the relationship: $s_b\, \bar c = -\, *\, s_d\, * \bar c$.
Using the duality symmetry transformation: $\bar c \to \pm \,i\, \dot c$, we note that we have: $s_b\, \bar c = \mp \, i\, *\, s_d\, \dot c$.
Now we explain the derivation of the transformation $s_d \, \dot c$ which is {\it not} present in the 
list of the co-BRST symmetry transformations that have been quoted in (7). In this context, we would like to mention that
we have the following from (7), namely;
\begin{eqnarray}
s_d \, c = -i\; \bigl [g\, (x\,p_y - y\, p_x) + p_z \bigr ] \equiv i\; \dot b,
\end{eqnarray}
where we have exploited the EL-EoM 
from the quantum version of the Lagrangian $L_b$ w.r.t. the gauge variable $\zeta$, which yields: 
$ \dot b = -\; \bigl [g\, (x\,p_y - y\, p_x) + p_z \bigr ]$.
Taking the first-order time derivative of  {\it both} the sides of the above equation, we find that 
\begin{eqnarray}
s_d \, \dot c = i\; \ddot b \equiv i\, b,
\end{eqnarray}
where we have used the EL-EoM: $ \ddot b = b$ that has been quoted in Eq. (10). Thus, we obtain, finally, the following 
expression from the above exercise
\begin{eqnarray}
s_b \, \bar c = \mp\,i\, *\; (i\, b) \equiv (\mp\, i) \; i\; (\pm\, i\, b) \equiv  i\; b,
\end{eqnarray}
where we have taken the discrete duality symmetry transformation: $b \to \pm\;i\;  b $. It is clear that we have obtained the {\it correct}
answer in the sense that the BRST transformation on $\bar c $ is equal to $i\, b$ 
(i.e. $s_b\, \bar c = i\; b$). As an alternative example, we now
concentrate on $s_d \, c = +\, *\; s_b\; *\, c \equiv \pm\; i\; *\, s_b\,\dot {\bar c}$ where we have taken the duality symmetry transformation
$c \to \pm\, i\, \dot {\bar c}$ [cf. Eq. (27)]. Ultimately, we obtain the following explicit transformation
\begin{eqnarray}
s_d \, c = \pm\, i\; *\; (i\;\dot  b) \equiv (\pm\, i)\; (i) \; (\mp\; i\; \dot b) \equiv i\, \dot b,
\end{eqnarray}
where we have exploited the strength of the discrete duality symmetry transformation: $\dot b \to \mp\, i\, \dot b $.
At this stage, we use the EL-EoM: $\dot b = \;- \, \bigl [g\, (x\, p_y - y\, p_x) + p_z \bigr ]$ 
(with respect to the gauge variable $\zeta$) which leads to the derivation of 
$s_d\, c = - \, i\, \bigl [g\, (x\, p_y - y\, p_x) + p_z \bigr ]$ that has been listed in the set of (anti-)co-BRST symmetry transformations (7).
In exactly similar manner, we can prove the validity of $s_{ab} = -\, *\, s_{ad}\, *$ as well as its counterpart: $s_{ad} = +\, *\, s_{ab}\, *$.
Thus, the above observations (in terms of the underlying continuous and discrete duality symmetry transformations of our 1D quantum
mechanical system of the FLPR model) provide the physical realizations of the mathematical relationship 
($\delta = \pm\, *\, d\,* $) between the (co-)exterior derivatives $(\delta)d$ of differential geometry [13-17] in the terminology of the 
off-shell nilpotent (anti-)BRST and (anti-)co-BRST symmetry transformations along with the set of discrete duality symmetry transformations (27).

We devote a bit of time now on establishing a {\it direct} connection between (i) the BRST and co-BRST symmetry transformations, and
(ii) the anti-co-BRST and anti-BRST symmetry transformations [that have been listed in equations (3) and (7)]. It is interesting
to point out that in this exercise, too, the set of discrete duality symmetry transformations (27) play a decisive role.  In fact,
under the {\it latter} symmetry transformations, we have the following relationship on the generic variables $\Phi_1$
and $\Phi_2$ of our theory that is described by the quantum version of the Lagrangian $L_b$. To be precise, for the algebraic
convenience, we obtain the following results after a couple of {\it successive} operations 
(see, e.g. [38]) of the discrete duality symmetry transformations
on the above generic variables, namely;
\begin{eqnarray}
&& *\, (*\, \Phi_1) = - \;\Phi_1,     \qquad \Phi_1 =\, x, \,y, \,z,  \,p_x, \, p_y, \, p_z, \, \zeta, \, b, \nonumber\\
 &&  *\, (*\, \Phi_2) = + \;\Phi_2,     \qquad \Phi_2 = \,c, \, \bar c,
\end{eqnarray}
which will be very useful in our goal to establish a {\it direct} connection between the nilpotent (anti-)BRST
and (anti-)co-BRST symmetry transformations. In fact, the signs on the r.h.s. of the above equation will play a {\it crucial}
role in our discussions that follow from here. To be precise, we observe that the 
following relationships\footnote{It will be noted that the role of the $*$ operation (i.e. the discrete duality symmetry transformations) 
on the square bracket of the l.h.s. of (34) is slightly {\it different} from the $*$ operations on the r.h.s. of the equation (28) where the
$*$ operation does {\it not} act {\it directly} on the symmetry operators $s_{(a)b}$ and $s_{(a)b}$.}
are {\it true}, namely;
\begin{eqnarray}
&& * \,\bigl [s_{(a)b} \,\Phi_1 \bigr ] = -\; \bigl (s_{(a)d} \, \Phi_1 \bigr ),  \qquad 
 *\, [s_{(a)d} \,\Phi_1] = +\;   \,\bigl  (s_{(a)b} \,\Phi_1 \bigr ), \nonumber\\
&& * \, \bigl [s_{(a)b}\,  \Phi_2 \bigr ] = +   \; \bigl (s_{(a)d} \,\Phi_2 \bigr ), \qquad
 *\, \bigl [s_{(a)d} \,\Phi_2 \bigr ] =  - \,  \bigl (s_{(a)b} \, \Phi_2 \bigr ), 
\end{eqnarray}
between the (anti-)BRST and (anti-)co-BRST symmetry transformations
where, in the case of $\Phi_2$,  we have to use EL-EoMs: 
$\ddot b = b, \; \dot b = -\; \bigl [g\, (x\,p_y - y\, p_x) + p_z \bigr ], \, \ddot c = c, \; \ddot {\bar c} = \bar c$. We 
take some explicit examples to establish the sanctity of the 
above relationships and their connection with the signs on the r.h.s. of (33). First of all, let us focus on the signs.
In the first {\it column} of the above equation (34), the signs
on the r.h.s. are governed by our observations in (33). On the other hand, these signs are {\it reversed} in the reciprocal relationships
in the second column. Let us take a couple of examples for the explanation of the above relations. As an example of the first entry (i.e. the l.h.s.)
in the above equation, we take $s_b\, x = -\; g\,y\, c$ where $\Phi_1$ is nothing but $x$. Applying the discrete duality 
(i.e. the $*$ symmetry)  transformations on its {\it both} the sides, we obtain the following  expression
\begin{eqnarray}
*\, s_b\, (*\, x) = - \;*\, g\, *\, y\, *\, c  \quad  \Longrightarrow \quad 
-\, s_d\, (\mp\, i\, x) = -\, (\pm\, i\, g)\, (\mp\,i\, y)\, (\pm\,i\, \dot {\bar c}),
\end{eqnarray}
where we have taken $*\, s_b = -\, s_d$ as is evident from the {\it first} entry of (34). It is clear from the above
equation that we have obtained $s_d \, x = -\, g\, y\, \dot {\bar c}$ which is the correct result. 
We lay emphasis on the fact  that the discrete duality  $*$ symmetry operation is applied on {\it all} the operators that are
present on the l.h.s. as well as on the r.h.s. of the BRST symmetry transformation: $s_b\, x = -\, g\, y\, c$. Now we consider 
the transformation: $ s_b\, \bar c = i \, b$ which is an example of the second entry of equation (34) where 
generic variable $\Phi_2$ has been chosen to be $\bar c $. Once again, applying the discrete duality
 symmetry transformations (i.e. the $*$ operation) on {\it both} the sides of $ s_b\, \bar c = i \, b$, we obtain the following explicit expression
\begin{eqnarray}
*\, s_b\, (*\, {\bar c}) = +\, i\, *\, ( b)  \quad  \Longrightarrow \quad 
+\, s_d\, (\pm\, i\, \dot c) =  i\, (\pm\,i\,b) \quad  \Longrightarrow \quad s_d \, \dot c = +\,i\, b,
\end{eqnarray}
where we have utilized: $*\, s_b = +\, s_d$ as is very clear from the {\it second} entry of (34). Taking a time derivative
of both the sides of the {\it last} entry of the above equation and using the EL-EoMs from $L_b$ as: $ \ddot c = c, \;
\dot b = -\; \bigl [g\, (x\,p_y - y\, p_x) + p_z \bigr ]$,
we obtain $s_d \, c = -i\, \bigl [g\, (x\,p_y - y\, p_x) + p_z \bigr ]$ which happens to be the {\it correct} result. In exactly
similar fashion, the sanctity of {\it all} the entries of equation (34) can be verified in their operator forms.

We wrap-up this section with the following remarks. First, the interplay between the continuous, off-shell nilpotent and absolutely anticommuting
(anti-)BRST and (anti-)co-BRST symmetry transformations [cf. Eqs. (3),(7)] and the discrete set of symmetry transformations [cf. Eq. (27)] provide
the physical realizations [cf. Eq. (28)] of the mathematical relationship (i.e. $\delta = \pm\, *\, d\, *$) that exists between the (co-)exterior
derivatives $(\delta)d$ of differential geometry [13-17].
 Second, it is {\it only} the set of discrete duality symmetry transformations (27) that is responsible
of the {\it direct} relationships between (i) the (co-)BRST symmetry transformations, and (ii)  the anti-BRST and anti-co-BRST symmetry transformations
[cf. Eq. (34)]. Third, we observe that the above relationships are {\it true} only on the on-shell conditions because we have to exploit the 
beauty and strength of the EL-EoMs: $\ddot b = b, \; \dot b = -\; \bigl [g\, (x\,p_y - y\, p_x) + p_z \bigr ], \; \ddot c = c, \; \ddot {\bar c} = \bar c$
in the proofs of some of the relationships that are listed in (28) and (34). Fourth, it should be noted that the $*$  operation (i.e. the discrete
duality symmetry transformations) in the equation (28) is applied only on the r.h.s. On the other hand, the $*$ operation is applied only on the 
l.h.s in the equation (34). Fifth, as far as the validity of the relationships in (34) is concerned, the {\it central} role is played by
the discreet duality symmetry transformations (27). The role of the continuous symmetry transformations, in some sense, is minor in the proof of the 
validity of relationships (34). For the verification as well as the explanation of (34), we have worked out some examples in (35) and (36). 
Finally, it is an undeniable truth that the existence of the
discrete duality symmetry transformations (27) is {\it crucial} in the proof the 1D quantum mechanical system of the FLPR model to be an
example of the Hodge theory.\\

\section{Physicality Criteria: Nilpotent Charges}

As far as the quantum version of the Lagrangian $L_b$ is concerned, right from the beginning, we note that
the ferminic (anti-)ghost variables are decoupled from the rest of the physical variables of our BRST-quantized theory.
In other words, there is {\it no} interaction (anywhere in the whole theory) between the (anti-)ghost variables and
the physical variables (carrying the ghost number equal to zero). Hence, the {\it total} qauntum Hilbert space of states (i.e. $|Hsqs>$)
is the {\it direct} product [39]
of the physical states (i.e. $|phys>$) and the ghost states (i.e. $|ghost>$). As a consequence, the
physical operators at the quantum level (carrying the ghost number equal to zero) act 
{\it only} on the physical states (i.e. $|phys>$). On the other
hand, the (anti-)ghost operators act {\it only} on the ghost states (i.e. $|ghost>$) 
at the quantum level. The result of the {\it latter}
operation would {\it always} be non-zero. However, within the framework of the BRST formalism, one of the key
requirements (of the properly BRST-quantized theory) is the physicality criteria where the physical states (i.e. $|phys>$)
are {\it those} states (in the $|Hsqs>$) that are annihilated by the  off-shell nilpotent and conserved 
(anti-)BRST charges. Thus, the operators (carrying the ghost number equal to zero) that are present in the expressions for the nilpotent and conserved
(anti-)BRST charges {\it must} annihilate the physical states. If the BRST-quantized theory happens to be a model for the
Hodge theory, then, the operator forms of the conserved and nilpotent (anti-)BRST and (anti-)co-BRST charges {\it must} annihilate
the physical states (due to the cohomological considerations w.r.t. {\it them}).

Against the backdrop of the above paragraph, we would like to elaborate on the reasons behind the requirement
of the physicality criteria where {\it all} the nilpotent charges of the models for the Hodge theory are invoked to
annihilate the physical states (existing in the total quantum Hilbert space of states). In this context, it is pertinent to point out that
{\it even} the physical states (i.e. $|phys>$) can be decomposed into a {\it unique} sum of the harmonic quantum states (i.e. $|hqs>$),
the BRST exact quantum states (i.e. $|\chi>$) and the BRST co-exact quantum states (i.e. $|\xi>$) due to the mathematical beauty and strength of the 
Hodge decomposition theorem\footnote{On a compact manifold without a boundary, any arbitrary $n$-form (i.e. $f_n$) can be written as a unique sum of 
the harmonic form ($h_n$) of degree $n$, an exact form (i.e. $d\, e_{n-1}$) and a co-exact form (i.e. $\delta\, c_{n+1}$) where
$e_{n-1}$ and $c_{n+1}$ are the non-zero and non-trivial forms of degrees $(n-1)$ and  $(n+1)$, respectively. In other words, we have:
$f_n = h_n + d\, e_{n-1} + \delta\, c_{n+1}$ where the operators $(\delta)d$ are the (co-)exterior derivatives of differential geometry
(see. e.g. [13-17] for details). The (co-)exterior derivatives
$(\delta)d$ are connected by a relationship: $\delta = \pm\, * \, d \, *$ where $*$ is the Hodge duality operator (that
is defined on the {\it above} compact manifold without a boundary). One of the key points (that is very useful in the 
context of any  arbitrary BRST-quantized model for the Hodge theory) is the observation that the harmonic form $h_n$ is 
{\it closed} and {\it co-closed} together (i.e. $d\, h_n = 0, \; \delta\, h_n = 0$) which is implied by the fact
that: $\Delta \, h_n = 0$ where $\Delta = (d + \delta)^2 \equiv \{ d, \, \delta\} $ is the Laplacian operator of the
set of three (i.e. $d, \delta, \Delta)$ de Rham cohomological operators of differential geometry
that are defined on the {\it above} compact manifold without a boundary.}. In other words, we have the following decomposition of the
physical state (i.e. $|phys>$), in the total quantum Hilbert space of states, namely;
\begin{eqnarray}
|phys> = |hqs> \,+\,  Q_b \; |\chi> \,+\,    Q_d \; |\xi>\; \equiv \; |hqs> \,+\,  Q_{ad} \; |\eta> \,+\,    Q_{ab} \; |\kappa>,
\end{eqnarray}
where the quantum states $|\chi>, \, |\xi>, \, |\eta>$ and $|\kappa>$ are the  non-trivial states in the
total quantum Hilbert space of states. The above equation has been written in {\it two} equivalent forms because of our
observations at the fag end of  our Sec. 4 where we have shown that, within the framework of the BRST formalism,
there are {\it two} quantum states that exist with the ghost numbers $(n+1)$ and $(n-1)$, respectively. The harmonic quantum states (i.e. $|hqs>$)
are the {\it most} symmetric states because (i) they are  the real physical states (i.e. $|Rphys>$) of our theory (which is an example
of the BRST-quantized Hodge theory), and (ii) they are annihilated by {\it all} the off-shell nilpotent and conserved 
(anti-)BRST as well as the (anti-)co-BRST charges of our theory. The latter 
conditions are implied by the requirement that the {\it unique} bosonic charge $Q_\omega$ annihilates 
(i.e. $Q_\omega\, |Rphys> \equiv Q_\omega\, |hqs> = 0 $) the most symmetric {\it real} physical states of our theory.

With the support of the background theoretical materials contained in
 the above paragraphs, we are at a stage where we provide the key reason behind
the requirement of the physicality criteria (i.e. $Q_{(a)b}\, |Rphys> \equiv Q_{(a)b}\, |hqs> = 0. \; Q_{(a)d}\, |Rphys> \equiv Q_{(a)d}\, |hqs> = 0 $)
w.r.t. the off-shell nilpotent and conserved (anti-)BRST and (anti-)co-BRST charges of our theory. In fact, this crucial requirement
leads the celebrated Dirac's quantization conditions (on the quantum versions of the gauge theories whose {\it classical}
counterparts are endowed with the first-class constraints). To be precise, the above quantizations conditions are nothing
but the requirement that the real physical states (i.e. $|Rphys> \equiv |hqs>$), in the total quantum Hilbert space of states, must be annihilated by
the operator forms of the constraints of the original {\it classical} gauge theory. In this context, first of all, we note that the first-order
{\it classical} Lagrangian [cf. Eq. (2)] is endowed with the primary constraint ($p_\zeta \approx 0$) and the secondary constraint
(i.e. $-\, [g\,(x\, p_y- y\, p_x) + p_z] \approx 0$). In the terminology of Dirac's prescription for the classification scheme of 
constraints [24,25], it has been shown, in our earlier work [32], that these constraints are of the first-class variety because they commute
with each-other. For the consistent quantum theory, corresponding to the classical gauge theory described by the Lagrangian $L_f$ [cf. Eq,. (2)], 
the Dirac quantization-condition requirements, on the real physical states (i.e. $|Rphys> \equiv |hqs>$)
of the BRST-quantized Hodge theory (with the primary and secondary constraints)  are:
\begin{eqnarray}
p_\zeta \, |Rphys> \equiv  p_\zeta \, |hqs> &=& 0, \nonumber\\
  -\, \bigl [g\,(x\, p_y- y\, p_x) + p_z \bigr ] \, |Rphys>  \equiv
-\, \bigl [g\,(x\, p_y- y\, p_x) + p_z \bigr ] \, |hqs> &=& 0.
\end{eqnarray}
These Dirac-quantization conditions are very sacrosanct and they must be satisfied within the framework of BRST formalism. It turns
out that, for our theory endowed with the first-class constraints, the above conditions emerge out when we demand that the real
physical states ($|Rphys>$) must be annihilated by the nilpotent charges of our theory, namely; 
\begin{eqnarray}
&& Q_{(a)b} \, |Rphys> \equiv  Q_{(a)b} \, |hqs> = 0  \; \Longrightarrow \; b\, |Rphys> = 0, \quad \dot  b\, |Rphys> = 0,  \nonumber\\
&& Q_{(a)d} \, |Rphys> \equiv  Q_{(a)d} \, |hqs> = 0  \; \Longrightarrow \; b\, |Rphys> = 0, \quad \dot  b\, |Rphys> = 0.
\end{eqnarray}
The above quantization conditions are implied by the requirement 
that $Q_\omega \, |Rphys> = 0$. These observations are consistent with the Hodge decomposition theorem
in the realm of differential geometry where the harmonic forms are {\it closed} and {\it co-closed} together. The quantization conditions
that have been obtained in (39) are exactly {\it same} as the requirement of the Dirac quantization conditions that have been quoted in (38)
for a {\it consistent} quantum theory.
We elaborate on {\it this} equivalence in the next paragraph.

In the context of the existence of the
operator forms of the first-class constraints (i.e. $p_\zeta \approx 0, \, -\, [g\,(x\, p_y- y\, p_x) + p_z] \approx 0$)
at the {\it quantum} level, it is pertinent to point out that these constraints have been traded with some other quantities at the level of the BRST-quantized theory that is described by the quantum version of the Lagrangian $L_b$ [cf. Eq. (1)]. For instance, it is straightforward to
note that the following are true, namely;
\begin{eqnarray}
p_\zeta = \frac{\partial L_b}{\partial \dot \zeta} = b, \qquad \quad \dot b =  -\, \bigl [g\,(x\, p_y- y\, p_x) + p_z \bigr ],  
\end{eqnarray}
where the {\it latter} entry in the above is nothing but the EL-EoM w.r.t. the ``gauge'' variable $\zeta$ from the Lagrangian $L_b$.
To be precise, we point out that the primary constraint (i.e. $p_\zeta \approx 0$) of the {\it classical} gauge theory is traded with the
Nakanishi-Lautrup type auxiliary variable $b$ and the secondary constraint (i.e. $-\, [g\,(x\, p_y- y\, p_x) + p_z] \approx 0$) is hidden
in the EL-EoM w.r.t. the variable $\zeta$ that is derived from the quantum version of the Lagrangian $L_b$ (because we observe that 
we have $\dot b =  -\,  [g\,(x\, p_y- y\, p_x) + p_z] $).
Thus, ultimately, our key claim is the fact
that the theoretical contents of (39) are {\it same} as (38) which are very sacrosanct according to Dirac's quantization conditions (for the 
consistent quantization of theories that are endowed with constraints at the {\it classical} level).

We end this section with the following remarks. First of all, we note that the quantization conditions (39), due to the requirement of the 
physicality criteria w.r.t. the off-shell nilpotent and conserved versions of the (anti-)BRST and (anti-)co-BRST charges, are the
{\it same} (i.e. $ b\, |Rphys> = 0, \; \dot  b\, |Rphys> = 0$) because of the fact that our quantum mechanical system is a model of
the one (0 + 1)-dimensional (1D)  gauge theory. Second, we have shown that, for the two (1 + 1)-dimensional (2D) 
massless and Stueckelberg-modified massive Abelian 1-form gauge theories (see, e.g. [8,40,41] for details), the physicality criteria w.r.t.
the conserved and nilpotent (anti-)BRST charges (i.e. $Q_{(a)b} \, |Rphys> \equiv  Q_{(a)b} \, |hqs> = 0$) leads to the aniihilation of the 
real physical states (i.e. $|Rphys> \equiv |hqs>$) by the operator form of the first-class constraints  that exist
at the {\it classical} level of these theories. On the 
other hand, the conserved and nilpotent (anti-)co-BRST charges in the physicality criteria (i.e. $Q_{(a)d} \, |Rphys> \equiv  Q_{(a)d} \, |hqs> = 0$)
lead to the annihilation of the real physical states (i.e. $|Rphys> \equiv |hqs>$) by the {\it dual} versions of the operator
forms the first-class constraints (of the {\it original} classical version of the above gauge theories [8,40,41]). This happens due the fact that
there is a set of discrete {\it duality} symmetries in these  theories where the {\it total}
gauge-fixing and kinetic terms of the BRST-quantized versions of the
Lagrangian densities are connected with one-another. Finally, we note that for our 1D quantum mechanical system of the FLPR model, there
exists a set of {\it duality} symmetry transformations (cf. Appendix B for details) where we have: $b \rightarrow \,\pm\, i\, b, \; 
\dot b \rightarrow \,\mp\, i\, \dot b$
which provide a beautiful connection between the off-shell nilpotent (anti-)BRST and (anti-)co-BRST  charges. This is the central reason behind our observation that the following physicality criteria
\begin{eqnarray}
Q_\omega \,|Rphys> = 0 \quad \Longrightarrow \quad 
Q_{(a)b} \; |Rphys> = 0  \quad \mbox{and} \quad  Q_{(a)d} \; |Rphys> = 0,  
\end{eqnarray}
lead to the quantizations conditions on the physcal states  that are exactly the {\it same} (i.e. $b\, |Rphys> = 0, \;\dot  b\, |Rphys> = 0$)
and they correspond to the annihilation of the real physical states (i.e. $|Rphys> \equiv |hqs>$)
of the quantum mechanical theory by the operator forms of the first-class constraints (because 
our present 1D quantum mechanical theory of FLPR model, within the framework of the BRST formalism,
turns out to be a tractable (and quite illustrative example) for the Hodge theory). \\

\section{Conclusions}

In our present endeavor, we have shown the existence of a set of {\it six} continuous symmetry transformations out of which {\it four} are  fermionic
(i.e. nilpotent) in nature and {\it two} are bosonic. In addition, we have demonstrated the presence of a couple of discrete {\it duality}
symmetry transformations which provide (i) the physical realizations of the Hodge duality $*$ operator of differential geometry [13-17], (ii) a deep
(and very interesting) connection between the off-shell nilpotent (anti-)BRST and (anti-)co-BRST symmetry transformations [cf. Eq. (28)] which turns
out to be the physical realizations of the mathematical relationship: $\delta = \pm \, *\, d \, *$ that  exists between the (co-)exterior derivatives
$(\delta)d$ of differential geometry (see, e.g. [13-17] for details), and (iii) a direct
connection between the (co-)BRST as well as the anti-BRST {\it and} anti-co-BRST symmetry transformations [cf. Eq. (34)]. 
All the above discrete and continuous symmetry transformations
(and corresponding conserved charges) {\it together} provide the physical realizations of the de Rham cohomological operators of differential geometry
at the algebraic level (see., e.g. our Appendix C for more details). We would like to lay emphasis, once again,  that the
existence of the complete set of
(i) the discrete symmetry transformations, and (ii) the continuous symmetry transformations (and corresponding conserved charges) 
are very crucial for our whole discussion.

The above continuous symmetries 
of our present model of Hodge theory are: (i) the infinitesimal and off-shell nilpotent (anti-)BRST symmetry transformations, (ii) the
off-shell nilpotent and infinitesimal (anti-)co-BRST symmetry transformations, (iii) the infinitesimal {\it unique} bosonic symmetry transformations,
(iv) the infinitesimal version of the ghost scale symmetry transformations, and (v) a set of a couple of discrete duality symmetry transformations
(for our 1D quantum mechanical FLPR model). These symmetries
 are characterized by very {\it specific} features. For instance, we observe that (i) the total kinetic terms of the Lagrangian
$L_b$ [cf. Eq. (1)] remain invariant under the (anti-)BRST symmetry transformations, (ii) the total gauge-fixing terms remain invariant under the
(anti-)co-BRST symmetry transformations, (iii) the total FP-ghost terms remain invariant under the 
{\it unique} bosonic symmetry transformations, (iv) only
the (anti-)ghost variables transform under the infinitesimal global (i.e. spacetime independent) ghost-scale symmetry transformations (and 
the {\it rest} of the physical variables do not transform at all), and (v) there is an electromagnetic duality type symmetry transformations
between the (anti-)BRST and (anti-)co-BRST conserved and nilpotent charges because we obseve that, under the discrete duality symmetry transformations,
we have: $Q_b \to Q_d, \; Q_d \to -\, Q_b, \; Q_{ab} \to Q_{ad}, \;  Q_{ad} \to -\, Q_{ab}$ which are just like
the transformations: ${\bf E} \to {\bf B}, \; {\bf B} \to -\, {\bf E} $ under which the source-free Maxwell's equations remain invariant
(where we have taken the natural unit: $c = 1$). In the above, the 3D electric and magnetic fields are denoted by the symbols
${\bf E}$ and ${\bf B}$, respectively.

Before we conclude this section, we would like to shed light on the symmetry operator relationships (28) and (34) where we have established
the connection(s) between the off-shell nilpotent (anti-)BRST and (anti-)co-BRST symmetry transformations with the help of the discrete
duality symmetry transformations in (27). As far as the symmetry operator relationships (28) are concerned,  we find that the {\it signs} on the r.h.s.
have been obtained by the trial and error method (where the close and careful observations have been 
taken into account). These operator relationships are {\it true} 
for the fermionic as we well as the bosonic variables $\Phi_1$ and $\Phi_2$ [cf. Eq. (33)] of our theory. On the other hand, the symmetry
operator relationships (34) provide the connection between (i) the BRST and co-BRST symmetry transformations, and (ii) the anti-BRST and
anti-co-BRST symmetry transformations where the {\it signs} on the r.h.s. are dictated by a couple of {\it successive} operations of the 
full set of discrete duality symmetry transformations (27) on the variables $\Phi_1$ and $\Phi_2$ of our theory. These have been
explained, in a clear fashion, by taking the examples of $\Phi_1$ and $\Phi_2$ separately and independently in (35) and (36). 
It is worthwhile
to lay emphasis on the fact that the discrete duality $*$ operation on the l.h.s. of (34) acts {\it separately} on (i) the symmetry transformation
operator, and (ii) the specifically chosen variable $\Phi_1$ or $\Phi_2$. This is why the square brackets have been used on the l.h.s.
On the other hand, the operations of the discrete duality
$*$ operations on the r.h.s. of (28) are very clear (because  they act, in succession, in a straightforward fashion on a given
variable $\Phi_1$ or $\Phi_2$ of our theory). It will be worthwhile to point out that, under the complete set of discrete duality $*$
symmetry transformations (27), the extended BRST algebra with the conserved ghost charge [cf. Eq. (23)] and the complete set of
the extended BRST algebra (C.6), are found to remain invariant (due to the full set of transformations (B.1) of the conserved 
charges under the complete set of  
discrete duality $*$ symmetry transformations).

We end our present section with the following concluding remarks. First of all, the 1D quantum mechanical system of FLPR model is very interesting
in itself because it represents an interesting example of a gauge theory where the {\it correct} results are obtained by 
taking into account the Gribov-type
gauge equivalent copies which are more general and somewhat {\it different} from the very important 
and useful suggestion made by Gribov himself in his seminar work [23]. Second,
if the beauty of a theoretically interesting and useful model is defined in terms of the numbers and varieties of symmetries it respects, our
present  1D quantum mechanical
system of FLPR model belongs to this {\it category} where we have shown the existence of a  set of {\it six} continuous symmetry transformations
in addition to {\it a couple} of discrete duality symmetry transformations which provide the physical realizations of the de Rham
cohomological operators of differential geometry (in the language of symmetries and conserved charges). Third, the selection of the spatial
two (0 + 2)-dimensional (2D)  general potential of the form $U(x^2 + y^2)$ is an {\it ingenious}  choice because it respects (i) the spatial 
2D rotational symmetry transformations, (ii) the parity symmetry transformations: $x \to -\, x,\; y \to -\, y$, and (iii)
all the {\it six} continuous and {\it a couple} of discrete duality symmetry transformations. Fourth, in  a couple of papers [42,43], the 
1D system of a particle on the torus and the 2D self-dual chiral bosonic field theory have been shown to be the models of Hodge theory. 
It will be nice to explore the possibilities of having more models of Hodge theory 
on the non-trivial topological objects like a torus (and/or the higher-genus Riemann surfaces)
and apply the supervariable/superfield approach to derive their nilpotent symmetries as has been done in [44]. Fifth, in a very recent 
publication [45], a unifying picture has been presented where the mathematical origin (see, e.g. [46] and references therein)
for the (anti-)BRST and (anti-)co-BRST has been emphasized and some {\it new} BRST-type symmetries have been pointed out. We would
like to devote time on the derivation of these {\it new} symmetries by applying the superfield/supervariable  approach.
Finally, it would be a very nice future endeavor [47]
to prove the field-theoretic and supersymmetric models of physical interest to be a set of tractable examples for the Hodge theory
where, by their very definition, many interesting continuous and discrete symmetries
would exist and such systems would provide a meeting-grounds for the mathematicians and 
theoretical physicists. \\

\vskip 0.8cm

\noindent
{\bf Acknowledgments}\\

\vskip 0.9cm

\noindent
We are thankful to Dr. Saurabh Gupta, Dr. Rohit Kumar and Dr. A. K. Rao for fruitful conversations and for taking interest in our present endeavor.
Both the authors dedicate, very humbly and respectfully, their present work to the memory of Prof. O. N. Srivastava (one of the top-class
experimental physicists of India) who passed away in the recent past due to Covid-Menace. Prof. Srivastava was one of the very influential 
well-wishers of our theoretical high energy physics group (at BHU, Varanasi) and  he always used to encourage us to apply the
sophisticated mathematics of BRST formalism to some interesting quantum mechanical systems of physical importance. It is our pleasure
to thank our {\it Reviewers} for their very useful and interesting comments.\\

\vskip 0.9 cm 
\begin{center}
{\bf Appendix A: \bf On the Bosonic Charge: Noether Theorem and Other Methods}\\
\end{center}

\vskip 0.9cm 

\noindent
The central purpose of this Appendix is to apply the celebrated Noether theorem,
due to our observations in (16),  to derive the conserved bosonic charge ($Q_\omega$) that turns out to be the generator for the
infinitesimal bosonic symmetry transformations  (13) and (14). The {\it latter} asymmetry transformations are generated by
a conserved charge ($Q_{\bar \omega}$) which is obtained from ($Q_\omega$) by taking into account a minus sign in front  of
it (i.e. $Q_{\bar \omega} = -\,Q_\omega $). This is 
due to our observation that: $Q_\omega + Q_{\bar \omega} = 0$. The stage is now set
to apply the Noether theorem, in its full blaze of glory [see, e.g. Eq. (16)], as follows:
\[
Q_\omega = (s_\omega \, x) \,\frac{\partial L_b}{\partial \dot x} + (s_\omega \, y) \,\frac{\partial L_b}{\partial \dot y}
+(s_\omega \, z)\, \frac{\partial L_b}{\partial \dot z} + (s_\omega \, \zeta)\, \frac{\partial L_b}{\partial \dot \zeta} + 
(s_\omega \, c) \,\frac{\partial L_b}{\partial \dot c} + (s_\omega \,\bar c)\, \frac{\partial L_b}{\partial \dot {\bar c}}
\]
\[ - \,\dot b\,  \bigl \{g\,(x\, p_y - y \,p_x) + p_z \bigr \} + b \, \frac{d}{dt}\, \bigl \{ g\,(x\, p_y - y \,p_x) + p_z \bigr \}.
\eqno (A.1)
\]
We note that the contributions of the {\it fifth} and {\it sixth} terms, on he r.h.s. of the above are zero because of the fact that we have: $ s_\omega c 
= 0 ,\; s_\omega \bar c = 0$. The explicit substitutions and careful determination of 
{\it all}  the existing terms, on the r.h.s. of (A.1),  lead to
\[
Q_\omega = b^2 -  \bigl [g\,(x\, p_y - y \,p_x) + p_z \bigr ]^2 \equiv b^2 - \dot b^2 \equiv b\, \ddot b - \dot b^2, 
\eqno (A.2)
\]
where we have used the EL-EoMs: $\ddot b = b, \; \dot b = -\, \bigl [g\,(x\, p_y - y \,p_x) + p_z \bigr ]$ derived from the
quantum version of the Lagrangian $L_b$ to obtain the simpler forms. It is straightforward to note that the above forms
of the charges are conserved (i.e. $\dot Q_\omega = 0$) if we exploit the beauty and strength of the 
appropriate forms of the EL-EoMs that are derived from $L_b$.

There are a few other simpler (and more appealing) methods to derive the exact expression for the 
{\it unique} bosonic charge $Q_\omega$ where we exploit the general relationships between the  continuous symmetry
transformations and their corresponding conserved Noether charges as {\it their} generators [cf. Eq. (6)]. For instance, let us
concentrate the following relationships:
\[
s_b Q_d = -\, i\, \{Q_d, \; Q_b \} = +\, i\, Q_\omega, \qquad \quad  s_d Q_b = -\, i\, \{Q_b, \; Q_d \} = +\, i\, Q_\omega.
\eqno (A.3)
\]
In the above, the l.h.s. of {\it both} the useful relationships can be computed explicitly by taking into account the
beauty and strength of the symmetry transformations (3) and (7) on the concise forms of the (anti-)BRST
and (anti-)co-BRST charges that have been listed in (5) and (9), respectively. It turns out that we have, for instance,  
the following:
\[
s_b Q_d = s_b \, \bigl [b \, \bar c - \dot b\, \dot {\bar c} \bigr ] \equiv i\, \bigl [ b^2 - \dot b^2 \bigr ] \equiv i\, Q_\omega.
\eqno (A.4)
\]
At this juncture, we have the freedom to use the 
theoretical strength of the appropriate EL-EoMs. For instance, the substitution of a suitable EL-EOM (e.g.
$\dot b = -\,  [g\,(x\, p_y - y \,p_x) + p_z  ]$) leads to the following, from the above simpler form (A.3), namely;
\[
  i\, Q_\omega = i\, \bigl [ b^2 - \{ g\,(x\, p_y - y \,p_x) + p_z \}^2 \bigr ] \quad \Longrightarrow  \quad
  Q_\omega = b^2 - \dot b^2 \equiv b\, \ddot b - \dot b^2,
\eqno (A.5)
\]
where we have used the EL-EoM: $\ddot b = b$ to obtain the precise expression  for the conserved Noether charge (A.2). In other words,
using the simple equations (A.3) and (A.4), we have been able to derive the conserved charge $Q_\omega$ that matches with the
expression for {\it it} (which has been derived by using the celebrated Noether theorem). In exactly similar fashion, it can be
seen that the explicit computation of $s_d Q_b$ [i.e. the l.h.s. of the {\it second} entry that is present in (A.3)] leads to the
derivation of the exact expression for the bosonic charge $Q_\omega$ that has already been quoted in (A.5) as well as in (A.2).

For the sake of the completeness of this Appendix, we now concentrate on the following anticommutators that correspond precisely to
the relationships between the continuous symmetry transformations and their generators as the conserved Noether charges, namely;
\[
s_{ad} Q_{ab} = -\, i\, \{Q_{ab}, \; Q_{ad} \} = - \, i\, Q_\omega, \qquad \quad  s_{ab} Q_{ad} = -\, i\, \{Q_{ad}, \; Q_{ab} \} = -\, i\, Q_\omega.
\eqno (A.6)
\]
We compute, first of all, the l.h.s. of the {\it first} entry in (A.6) which is explicitly written as:
\[
s_{ad} Q_{ab} = s_{ad} \, \bigl [ b\, \bar {\dot c} - \dot b \, \bar c] \equiv i\, b\, \frac{d}{dt}
\Bigl [ g\,(x\, p_y - y \,p_x) + p_z \Bigr ] - i\, \dot b \, \Bigl [ g\,(x \, p_y - y \, p_x) + p_z  \Bigr ].
\eqno (A.7)
\]
At this juncture, we have the freedom to use the equations of motion. For instance, using the 
EL-EoMs: $\ddot b = b, \, \dot b = -\,  [g\,(x\, p_y - y \,p_x) + p_z]$, we obtain the following 
different (but equivalent) forms of the expressions for
the bosonic charge $Q_\omega$, namely;
\[
Q_\omega =  b\, \ddot b - \, \dot b^2 \equiv b^2 + \dot b \,\bigl [g\,(x\, p_y - y \,p_x) + p_z \bigr ],
\eqno (A.8)
\]
where we have used the theoretical strength of the relationships (A.6), too. In other words, we have obtained, ultimately, the 
precise expression for the bosonic charge $Q_\omega$ that has been derived by the straightforward 
application of the Noether theorem in (A.2). Exactly
similar kinds of computations can be performed, as far as  the {\it second} entry of (A.6) 
is concerned, to obtain the precise expression
for the conserved (but non-nilpotent) version of the {\it unique} bosonic charge $Q_\omega$ of our present theory.

We end our present Appendix with the {\it final} remarks that we have exploited the
theoretical beauty and strength of 
(i) the celebrated Noether theorem, and (ii) the standard (as well as the general) relationship between the
continuous symmetry transformations and their generators as the Noether conserved charges, to obtain the precise 
expression\footnote{One can obtain the precise expression for the bosonic charge $Q_\omega$ as:
$Q_\omega = +\, \{ Q_{b}, \; Q_{d} \} = -\, \{ Q_{ad}, \; Q_{ab} \}$ provided one chooses
the factor $(-\,i)$ instead of the factor $(+\,i)$ in the definitions of the bosonic symmetry
transformations (13) and (14), respectively. It is very clear that the {\it former} choice of the $i$ factor [i.e. $(-\,i)$]
would lead to the derivation of the Noether charge with a minus sign in (A.2) which will, ultimately, imply
that we have the precise expression: $Q_\omega = +\, \{ Q_{b}, \; Q_{d} \} = -\, \{ Q_{ad}, \; Q_{ab} \}$ for our theory.}
for the bosonic charge $Q_\omega$ (i.e. $Q_\omega = \{ Q_{ad}, \; Q_{ab} \} = -\, \{ Q_{b}, \; Q_{d} \} $)
which matches with the standard conserved bosonic charge $Q_\omega$ that has been precisely derived in (A.2)
by exploiting the theoretical beauty (as well as the mathematical strength) of the Noether theorem.\\

\vskip 0.9cm 
\begin{center}
{\bf Appendix B: \bf On the Duality Symmetry Invariance: Algebra (23)}\\
\end{center}

\vskip 0.9cm

\noindent
The central purpose of this Appendix is to establish that the algebraic structure (23) between the conserved
ghost charge and 
the {\it rest} of the other charges of our theory remains invariant under a set of discrete duality symmetry transformations.
To this goal in mind, we elaborate here a few relevant theoretical stuffs that are very essential.

As pointed out earlier, in the {\it footnote} after the equation (21), it is very straightforward to note that the FP-ghost
terms (i.e. $- i \, \dot {\bar c}\, \dot c -\, i\,\bar c\; c $) 
of the quantum version of the Lagrangian $L_b$ remain invariant under the discrete symmetry
transformations in the ghost-sector: $ c \to \pm\,i\, \dot {\bar c}, \;  {\bar c} \to \pm\,i\, \dot { c},
 \; \dot c \to \mp\, i\,  {\bar c}, \;  \dot {\bar c} \to \mp\,i\,  { c}$. However, we find that the expression
for the ghost charge $Q_g$ [cf. Eq. (21)] undergoes the following transformation under the {\it above} discrete symmetry transformations: 
$ Q_g \to -\; Q_g$. We note further that the concise forms of (i) the (anti-)BRST charges [cf. Eq. (5)], 
(ii) the (anti-)co-BRST charges [cf. Eq. (9)], (iii)  the bosonic charge [cf. Eq. (18)], and (iv) the ghost charge [cf. Eq. (21)]
{\it together} transform as
\[
Q_b \longrightarrow Q_d, \qquad Q_{ab} \longrightarrow Q_{ad}, \qquad Q_d \longrightarrow -\; Q_b,
\]
\[
Q_{ad} \rightarrow -\; Q_{ab}, \qquad Q_{g} \rightarrow -\; Q_{g}, \qquad Q_\omega \rightarrow -\; Q_\omega, 
\eqno (B.1)
\]
under the following {\it complete} set of the discrete duality symmetry transformations:
\[
c \longrightarrow \;\pm\,i\, \dot {\bar c}, \qquad   {\bar c} \longrightarrow \;\pm\,i\, \dot { c}, \qquad
 \; \dot c \longrightarrow \;\mp\, i\,  {\bar c}, \qquad   \dot {\bar c} \longrightarrow \;\mp\,i\,  {c},
\]
\[
~~~~~~~~~~~~~~~~~~~~~~~~~~~~~~~~~~~~b \longrightarrow \;\pm \;i\; b, \qquad {\dot b} \longrightarrow \;\mp \;i\; {\dot b}.~~~~~~~~~~~~~~~~~~~~~~~~~~~~~~~
\eqno (B.2)
\]
It is very interesting to point out that the transformations in (B.1) leave the algebraic structure of the
equation (23) intact. In other words, the algebra between the conserved ghost charge and the {\it rest} of the other
conserved charges of our theory remains invariant under the discrete duality symmetry transformations (B.2) [at the
level of the conserved  charges (B.1) which are generated due to the discrete symmetry transformations (B.2)]. 
It is worthwhile to point out that the discrete duality symmetry transformations (B.1) are a part of the {\it total}
discrete duality symmetry transformations that have been listed in (27). We would like to point out that $\dot b \to \mp\, i\, \dot b$
is consistent with the individual discrete duality symmetry transformations on all the variables that are present in
the EL-EoM: $\dot b = -\,  [g\,(x\, p_y - y \,p_x) + p_z]$ that is derived from $L_b$ w.r.t. the gauge variable $\zeta (t)$.

We end this Appendix with the following crucial remarks. First of all, we would like to point out that {\it all} the variables and their
first-order time derivatives have been treated {\it independently} because of the fact that we are dealing with the BRST-quantized 
quantum mechanical system within the framework of the (anti-)BRST and (anti-)co-BRST invariant {\it Lagrangian} formulation (where, in general,
the coordinates and their first-order time derivatives are always treated on independent footings). This is the precise reason behind the
independent (and different) transformations for the pairs of quantities [e.g. $(b, \; \dot b), \; (c, \; \dot c), \; (\bar c, \; \dot {\bar c})$]
in our complete set of transformations in (B.2) as well as in (27). Second, let us concentrate on the non-trivial examples of some of the 
algebraic structures [cf. Eq. (23)] and explicitly check their transformation 
properties under (B.1) to establish that the algebra (23) remains invariant under the complete  set of transformations 
for the full set of conserved charges of our theory, namely;
\[
i\, \bigl [ Q_g, \; Q_b \bigr ] = +\; Q_b \quad \Longrightarrow  \quad  i\, \bigl [ -\, Q_g, \; Q_d \bigr ] = +\; Q_d \quad
\Longrightarrow \quad i\, \bigl [ Q_g, \; Q_d \bigr ] = -\; Q_d,
\]
\[
i\, \bigl [ Q_g, \; Q_{ab} \bigr ]  = -\; Q_{ab} \quad \Longrightarrow  \;  i\, \bigl [ -\, Q_g, \; Q_{ad} \bigr ] = -\; Q_{ad} \quad
\Longrightarrow \; i\, \bigl [ Q_g, \; Q_{ad}\bigr ] = +\; Q_{ad}, 
\]
\[
i\, \bigl [ Q_g, \; Q_d \bigr ]  = -\; Q_d \quad \Longrightarrow  \;  i\, \bigl [ -\, Q_g, \; -\,  Q_b \bigr ] = +\; Q_b \quad
\Longrightarrow \quad i\, \bigl [ Q_g, \; Q_b \bigr ] = +\; Q_b,
\]
\[
i\, \bigl [Q_{g}, \; Q_{ad} \bigr ]  = +\; Q_{ad} \; \Longrightarrow  \quad  i\, \bigl [ -\, Q_g, \; -\, Q_{ab} \bigr ] = -\; Q_{ab} \;
\Longrightarrow \quad i\, \bigl [ Q_g, \; Q_{ab}\bigr ] = -\; Q_{ab}.
\eqno (B.3)
\]
It goes without saying that the {\it rest} of the commutators of (23) are trivially 
invariant under the transformations (B.1). The above observations in (B.3)
demonstrate {\it explicitly} that the algebraic structure (23) remains invariant under the 
discrete symmetry transformations in (B.2). Finally,
we christen the above transformations [of the conserved charges in (B.1)] as the discrete ``duality'' symmetry transformations because the
transformations between the conserved (anti-)BRST charges ($Q_{(a)b}$) and the conserved 
(anti-)co-BRST charges ($Q_{(a)d}$) are similar to the
electric-magnetic duality (i.e. ${\bf E} \to {\bf B}, \; {\bf B} \to -\,{\bf E}$)
symmetry transformations that {\it exist} for the {\it simple} source-free Maxwell's theory of the
classical electromagnetism where the symbols (${\bf E},\, {\bf B}$) stand for the electric and magnetic fields. \\

\vskip 0.5cm
 
\begin{center}
{\bf Appendix C: \bf On the Cohomological Operators, Symmetries and Charges}\\
\end{center}

\vskip 0.5cm

\noindent
We have observed that the {\it four} fermionic (i.e. off-shell nilpotent) continuous and infinitesimal (anti-)BRST
and (anti-)co-BRST symmetry transformations [cf. Eqs (3),(7)] and a {\it unique} set of bosonic symmetry transformations [cf., e.g. Eqs. (13),(14)]
provide the physical realizations of the de Rham cohomoloigial operators ($d, \, \delta, \, \Delta $) of differential geometry
in the language of symmetry operators (and corresponding conserved charges). The central theme of our present Appendix is to demonstrate that
there is two-to-one mapping between the symmetry operators and conserved charges of our BRST-quantized theory on one hand and 
the cohomological operators on the other hand (as far as their algebraic structures are concerned). In this connection, it is pertinent
to point out that the theoretical and mathematical contents of our Secs. 2-5 play the decisive roles
 in this verification.

It is an interesting 
and very useful observation of our present investigation that the following well-known Hodge algebra amongst the de Rham cohomological operators
($d, \; \delta, \; \Delta$)
of differential geometry (see, e.g. [13-17] for details)
\[
d^2 = 0, \qquad \delta^2 = 0, \qquad \{d, \; \delta\} =
\Delta=  (d + \delta)^2, 
\]
\[
[\Delta, \;  {d} ] = 0, \qquad [\Delta, \; \delta ] = 0, \qquad d\, \delta + \delta\, d \neq 0, \eqno (C.1)
\]
are realized in terms of the specific set of infinitesimal continuous symmetry transformation operators ($s_b, \; s_d, \; s_\omega$) of our
theory, namely;
\[
s_b^2 = 0, \qquad s_d^2 = 0, \qquad \{s_b, \; s_d\} =
s_\omega =  (s_b + s_d)^2, 
\]
\[
[s_\omega, \;  s_{b} ] = 0, \qquad [s_\omega, \; s_d ] = 0, \qquad s_b\, s_d + s_d\, s_b \neq 0. \eqno (C.2)
\]
Ir should be noted that the Laplacian operator appears to be like a Casimir operator for the whole algebra that is satisfied by the
de Rham cohomological operators of differential geometry [cf. Eq. (C.1)]. However, this operator is {\it not} like the Casimir operator(s) of the
Lie algebra(s). Similar statement can be made for the infinitesimal bosonic symmetry transformation $s_\omega$ in the 
context of the algebra (C.2). It is worthwhile to mention, at this juncture, that another set of symmetry operators (of our
BRST-quantized theory of the FLPR model) can {\it also} provide the physical realization of the Hodge algebra in (C.1). This set consists
of the symmetry operators ($s_{ad}, \; s_{ab}, \; s_{\bar \omega} = -\, s_\omega$) of our theory
because we observe that the
following useful algebra is satisfied by these operators
\[
s_{ad}^2 = 0, \qquad s_{ab}^2 = 0, \qquad \{s_{ad}, \; s_{ab}\} =
s_{\bar \omega} =  (s_{ad} + s_{ab})^2, \]
\[ [s_{\bar \omega}, \;  s_{ad} ] = 0, \qquad [s_{\bar \omega}, \; s_{ab} ] = 0, \qquad s_{ad}\, s_{ab} + s_{ab}\, s_{ad} \neq 0, \eqno (C.3)
\]
which is just like the algebra of the de Rham cohomological operators of differential geometry that is quoted in (C.1). Thus, it
is evident that, at the algebraic level, we have a two-to-one mapping between the symmetry transformation operators and the set of 
{\it three} geometrical de Rham cohomological operators of differential geometry as:
\[
(s_b, \; s_{ad}) \Longrightarrow d, \qquad  (s_{ab}, \; s_{ad}) \Longrightarrow \delta, \qquad 
(s_\omega, \; s_{\bar \omega}) \Longrightarrow \Delta. \eqno (C.4)
\]
Hence, our present 1D quantum mechanical system of the FLPR model is an example for the Hodge theory and its various symmetries 
have connection with the de Rham cohomological operators of differential geometry as  far as the algebraic 
{\it structure} is concerned.

We would like to lay emphasis on the existence as well as importance of (i) the infinitesimal version of the ghost-scale
symmetry transformations, and (ii) the beauty as well as the importance  of the discrete duality symmetry transformations
(of our present 1D quantum mechanical system of FLPR model) in the context of the analogue of the Hodge algebra
(which is satisfied by the operator forms of
symmetries and conserved charges of our present model). First of all, we note that the equation
(25) of our Sec. 4 demonstrates that the ghost number increases by {\it one} when we apply a BRST charge [or an anti-co-BRST charge]
in its operator form on the state $|\psi>_n$
with the ghost number $n$ (i.e. $i\, Q_g \, Q_b \, |\psi>_n = (n + 1)\, Q_b \, |\psi>_n, \;
[i\, Q_g \, Q_{ad} \, |\psi>_n = (n + 1)\,  Q_{ad} \, |\psi>_n]$). This observation is {\it exactly} same as the operation
of an exterior derivative $d$ (i.e. $d = dx^\mu \, \partial_\mu, \; d^2 = 0$) on a differential form $ f_n$
of degree $n$ (i.e. $d\; f_n  \sim f_{n+1}$). On the other hand, we observe that
the relationships:  $i\, Q_g \, Q_d \, |\psi>_n = (n - 1)\, Q_d \, |\psi>_n, \; 
[i\, Q_g \, Q_{ab} \, |\psi>_n = (n - 1)\, Q_{ab} \, |\psi>_n]$ demonstrate that the co-BRST charge [or the anti-BRST charge] decreases the ghost 
number by {\it one} when it acts on a state with the ghost number equal to $n$ (i.e. $|\psi>_n$). This observation is exactly same as the 
operation of a co-exterior derivative $\delta$
(i.e. $\delta = \pm\, *\, d\, *, \; \delta^2 = 0$) on a differential form of degree $n$ (i.e. $ f_n$).
In other words, we observe that $\delta \, f_n \sim f_{n - 1}$. Finally, we note that the operation of the {\it unique} bosonic
charge $ Q_\omega = -\, Q_{\bar \omega}$ on the quantum state of the ghost number $n$ (i.e. $|\psi>_n$) does {\it not}
change its ghost number (i.e. $ i\, Q_g \, Q_{\omega} \, |\psi>_n = (n + 0)\, Q_{\omega} \, |\psi>_n)$ which is exactly similar to the
operation of the Laplacian operator $\Delta$ on a differential form of degree $n$ (i.e. $ \Delta\, f_n \sim f_n$). Hence,
 the infinitesimal ghost-scale symmetry transformations and corresponding conserved ghost charge play a crucial role in 
corroborating (and bolstering) the following mapping
\[
(Q_b, \; Q_{d}, \; Q_\omega) \Longrightarrow (d,\; \delta, \; \Delta), \qquad \mbox{and} \qquad 
(Q_{ad}, \; Q_{ab}, \; Q_{\bar \omega} \equiv -\, Q_\omega) \Longrightarrow (d,\; \delta, \; \Delta), \eqno (C.5)
\]
which demonstrates that the ghost-scale symmetry transformations and their generator 
(i.e. the conserved ghost charge) are very important as they capture 
a crucial feature of the de Rham cohomological operators of differential geometry (as far as their operations 
 on the differential form  of a given degree and ensuing consequences are concerned).

We dwell a bit, at this stage, on the importance and usefulness of the discrete duality symmetry transformations 
[cf. Eq. (27)] in our theory. In particular, we try to highlight their importance in the context of (i) proving the analogue
of the mathematical relationship: $\delta = \pm\, * \, d \, *$ in the language of the interplay between
the continuous and discrete symmetry transformations, (ii)
the invariance of the algebraic structure of the extended BRST algebra, and  (iii) providing
a {\it direct} connection between  the off-shell nilpotent (anti-)BRST and (anti-)co-BRST symmetry transformations.
First of all, we note that we have already obtained a couple of relationships 
between the (co-)BRST symmetry transformations $(s_d)s_b$ as: $s_d  = \, +\, * \, s_b \, *$ and $s_b  = \, -\, * \, s_d \, *$
in our Sec. 5 where, in the specific cases, the equations of motions $\ddot b = b, \; \dot b = - \, [g \, (x\, p_y - y\, p_x) + p_z], 
\;\ddot c = c, \; \ddot {\bar c} = {\bar c}$
have been used. In this context, mention can be made of 
the relationships: $s_d \, c = +\, *\, s_b \, *\, c$ and  $s_b \, \bar c = -\, *\, s_d \, *\, \bar c$
where the above equations of motions have been utilized. In the proof of the relationships 
(e.g. $s_d \, c = +\, *\, s_b \, *\, c, \;s_b \, \bar c = -\, *\, s_d \, *\, \bar c$) the
continuous and discrete symmetries play a decisive role {\it together}
where $*$ corresponds to the discrete duality symmetry transformations that have been listed in (27).
Thus, we have been able to provide  the physical realizations of the mathematical relationship: $\delta = \pm\, * \, d \, *$
in the language of the interplay between the continuous and discrete 
duality symmetry transformations of our theory.

A close look at the Secs. 2, 3 and 4 as well as the Appendices A and B, establishes 
the following extended version of the BRST algebra that is constituted by the {\it four} fermionic 
[i.e. off-shell nilpotent and absolutely anticommuting (anti-)BRST and (anti-)co-BRST] charges and a couple  (i.e. $Q_\omega, \; Q_g$)
of  bosonic conserved charges, namely;
\[
 Q_{(a)b}^2 = 0, \;\;\quad   Q_{(a)d}^2 = 0, \;\;\quad  \{Q_b, \; Q_{ab} \} = 0, \;\;
\quad \{Q_d, \; Q_{ad} \} = 0,\;\; \quad \{Q_d, \; Q_{ab} \} = 0, \]
\[ \{Q_b, \; Q_{ad} \} = 0, \qquad \quad Q_\omega = -\, \{Q_b, \; Q_d \} \equiv  \{Q_{ab}, \; Q_{ad} \}, \qquad \quad i\, [Q_g, \; Q_b ] = +\; Q_b,
\]
\[ [Q_\omega, \; Q_r] = 0, \,\quad r = b, ab, d, ad, g, \,\omega, \quad i\, [Q_g, \; Q_d ] = -\; Q_d, \quad i\, [Q_g, \; Q_{ab} ] = -\; Q_{ab},
\]
\[
 i\, [Q_g, \; Q_{ad} ] = +\; Q_{ad}, \;\;\qquad
 i\, [Q_g, \; Q_g ] = 0, 
\;\; \qquad i\, [Q_g, \; Q_\omega ] = 0.  ~~~~~~~~~~~~~~~~~~\eqno (C.6)
\]
It is straightforward to note that the {\it all} the conserved charges of the above extended BRST algebra, transform under the
discrete duality symmetry transformations (27), as illustrated and clearly written in (B.1). It is illuminating to point out that 
the above algebra (C.6) remains invariant under the transformations of the charges in (B.1) which have been derived from the {\it basic}
discrete duality symmetry transformations (B.2) that is a part of (27).

The full set of the discrete duality symmetry transformations also play a {\it crucial} role in proving a {\it direct} 
connection [cf. Eq. (34)] between (i) the off-shell nilpotent BRST and co-BRST symmetry transformations, and (ii) the fermionic anti-BRST
and anti-co-BRST symmetry transformations. To pinpoint precisely, the above direct connection (34) is established where the discrete
duality symmetry transformations (27) play a {\it pivotal} role (and the continuous nilpotent symmetry transformations are used
for {\it minor} purposes). We would like to lay emphasis on the fact that, in the proof of (28) and (34), a few EL-EoMs:
$\ddot b = b, \; \dot b = - \, [g \, (x\, p_y - y\, p_x) + p_z], \;\ddot c = c, \; \ddot {\bar c} = {\bar c}$,
derived from the quantum version of the Lagrangian $L_b$, are utilized. Thus, the relationships in (28) and (34)
are valid on the {\it on-shell} only where some of the EL-EoMs play a decisive role in their proof.\\

\vskip 0.3cm

\noindent
{\bf Conflicts of Interest}\\

\vskip 0.3cm

\noindent
The authors declare that there are no conflicts of interest. \\

\vskip 0.3cm

\noindent
{\bf Data Availability}\\

\vskip 0.3cm

\noindent
No data were used to support this study.\\


\vskip 0.5cm


\begin{thebibliography}{99}  
\bibitem{RPM1}     M. B. Green, J. H. Schwarz, E. Witten, Superstring Theory, Vols. 1 and 2, \\
                   Cambridge University Press, Cambridge (1987).
\bibitem{RPM2}     J. Polchinski, String Theory, Vols. 1 and 2,\\
                   Cambridge University Press, Cambridge (1998).   
\bibitem{RPM3}     D. Lust, S. Theisen, Lectures in String Theory, Springer-Verlag, New York (1989). 
\bibitem{RPM4}     K. Becker,  M. Becker, J. H. Schwarz, String Theory and M-Theory,\\
                   Cambridge University Press, Cambridge (2007).  
\bibitem{RPM5}     D. Rickles, A Brief History of String Theory From Dual Models to M-Theory,\\ Springer, Germany  (2014). 
\bibitem{RPM6}     R. Kumar, S. Krishna, A. Shukla, R. P. Malik,\\
                   {\it Int. J. Mod. Phys.} A {\bf 29} (2014) 1450135. (a brief review)
\bibitem{RPM7}     S. Krishna, R. Kumar, R. P. Malik, {\it Annals of Physics} {\bf 414} (2020) 168087.                  
\bibitem{RPM8}     B. Chauhan, S. Kumar, A. Tripathi, R. P. Malik,\\
                  {\it Advances in High Energy Physics} {\bf 2020} (2020) 3495168.
\bibitem{RPM9}      R. Kumar, R. P. Malik, {\it Eur. Phys. J.} {\bf C 73} (2013) 2514.
\bibitem{RPM10}    S. Krishna, R. P. Malik, {\it Euro. Phys. Lett.} {\bf 109} (2015) 31001.    
\bibitem{RPM11}    S. Krishna, R. P. Malik, {\sl Annals of Physics} {\bf 355} (2015) 204.
\bibitem{RPM12}    Saurabh Gupta, R. P. Malik, {\it Eur. Phys. J.} C {\bf 68} (2010) 325.
\bibitem {RPM13}    T. Eguchi, P. B. Gilkey, A. Hanson, {\it Physics Reports} {\bf 66} (1980) 213.
\bibitem {RPM14}    S. Mukhi, N.  Mukunda, Introduction to Topology, Differential  Geometry 
                    and Group Theory for Physicists, Wiley Eastern Private  Limited, New Delhi (1990).    
\bibitem{RPM15} 	K. Nishijima, {\it Prog. Theor. Phys.}  80 (1988) 897.   
\bibitem{RPM16}     J. W. van Holten, {\it Phys. Rev. Lett.} {\bf 64} (1990) 2863.    
\bibitem {RPM17}    M. G{\"o}ckeler, T. Sch{\"u}cker,  Differential Geometry, Gauge Theories   
                    and Gravity, Cambridge University Press, Cambridge 1987.
\bibitem{RPM18}     C. Becchi, A. Rouet, R. Stora, {\it  Phys. Lett.} B {\bf 52} (1974)  344.    
\bibitem{RPM19}     C. Becchi, A. Rouet, R. Stora, {\it  Comm. Math. Phys.} {\bf 42}  (1975) 127.  
\bibitem{RPM20}     C. Becchi, A. Rouet, R. Stora,  {\it Annals of  Physics} {\bf 98} (1976)  287.   
\bibitem{RPM21}     I. V. Tyutin {\it Lebedev Institute Preprint}, Report Number: {\bf FIAN-39} (1975) \\(unpublished),
                    {\bf arXiv:0812.0580 [hep-th]}.                     
\bibitem{RPM22}     R. Friedberg, T. D. Lee, Y. Pang, H. C. Ren, {\it Annals of Physics} {\bf 246} (1996) 381.
\bibitem{RPM23}     V. N. Gribov, {\it Nucl. Phys.} B {\bf 139} (1978) 1. 
\bibitem{RPM24}     P. A. M. Dirac, {\it Can. J. Math.} {\bf 2} (1950) 129; Lectures on Quantum Mechanics (Belfer Graduate
                    School of Science), Yeshiva University Press, New York (1964).
\bibitem{RPM25}     K. Sundermeyer, {\it Constrained Dynamics: Lectures Notes in Physics}, Vol. 169 \\
                    Springer-Verlag, Berlin (1982).
\bibitem{RPM26}     K. Fujikawa, {\it Nucl. Phys.} B {\bf 468} (1996) 355.
\bibitem{RPM27}     V. M. Villanueva, J. Govaerts, J. -L. Lucio-Martinez, {\it J. Phys.} A {\bf 33} (2000) 4183.
\bibitem{RPM28}     J. Govaerts, J. R. Klauder, {\it Annals of  Physics} {\bf 274} (1999) 251. 
\bibitem{RPM29}     A. S. Nair, Saurabh Gupta, {\it Mod. Phys. Lett.} A {\bf 39} (2024) 2350186.
\bibitem{RPM30}     L. Faddeev, R. Jackiw, {\it Phys. Rev. Lett.} {\bf 60} (1988) 1692.
\bibitem{RPM31}     J. Barcelos-Neto, C. Wotzasek, {\it Int. J. Mod. Phys.} A {\bf 7} (1992) 4981.
\bibitem{RPM32}     R. P. Malik, {\it Euro. Phys. Lett.} {\bf 144} (2023) 42002.
\bibitem{RPM33}     A. S. Nair, R. Kumar, Saurabh Gupta, {\it Eur. Phys. J. Plus} {\bf 138} (2023) 1107. 
\bibitem{RPM34}     D. Nemeschansky, C. Preitschopf, M. Weinstein, {\it Ann. Phys.} {\bf 183} (1988) 226.
\bibitem{RPM35}     H. Ruegg, M. Ruiz-Altaba, {\it Int. J. Mod. Phys.} A {\bf 19} (2004) 3265.
\bibitem{RPM36}     A. K. Rao, R. P. Malik, {\it Euro. Phys. Lett.} {\bf 135} (2021) 21001. 
\bibitem{RPM37}     F. G. Scholtz, S. V. Shabanov, {\it Annals of  Physics} {\bf 263} (1999) 119. 
\bibitem{RPM38}     S. Deser, A. Gomberoff, M. Henneaux, C. Teitelboim, {\it Phys. Lett.} B {\bf 400} (1997) 80. 
\bibitem{RPM39}     T. Kugo, I. Ojima,  {\it Prog. Theor. Phys.} (Suppl.) {\bf 66}  (1979) 1. 
\bibitem{RPM40}     R. P. Malik, {\it J. Phys. A: Math. Gen.} {\bf 34} (2001) 4167.
\bibitem{RPM41}     T. Bhanja, D. Shukla, R. P. Malik, {\it Eur. Phys. J.} {\bf C 73 } (2013) 2535.
\bibitem{RPM42}     V. K. Pandey, B. P Mandal, {\it Advances in High Energy Physics} {\bf 2017} (2017) 6124189.
\bibitem{RPM43}     S. Upadhyay, B. P. Mandal, {\it Eur. Phys. J.} C {\bf 71} (2011) 1759.
\bibitem{RPM44}     R. Kumar, {\it Euro. Phys. Lett.} {\bf 106} (2014) 51001.
\bibitem{RPM45}     B. P. Mandal, S. K. Rai, R. Thibes, {\it Euro. Phys. Lett.} {\bf 144} (2023) 14001.
\bibitem{RPM46}     S. Krishna, A. Shukla, R. P. Malik, {\it Mod. Phys. Lett.} {\bf 26} (2011) 2739.
\bibitem{RPM47}     R. P. Malik, {\it etal.}, in preparation.
\end{thebibliography}
\end{document}